\documentclass{emulateapj}
\usepackage{natbib}
\usepackage{bm}
\usepackage{tabu}
\usepackage{mathrsfs,amsmath}
\usepackage{mathtools}
\usepackage{graphicx}
\usepackage{epstopdf}
\usepackage[english]{babel}
\usepackage{xcolor}
\usepackage[colorlinks,linkcolor={blue},citecolor={blue},urlcolor={blue}]{hyperref}

\defcitealias{kopp18b}{K18}
\defcitealias{r10}{R10}

\shorttitle{Dynamical Relics of the Galactic Halo}
\shortauthors{Yuan et al.}

\begin{document}

\title{Dynamical Relics of the Ancient Galactic Halo} 

\author{Zhen Yuan\altaffilmark{1}, G.~C. Myeong\altaffilmark{2,3}, Timothy C. Beers\altaffilmark{4,5}, N. W. Evans\altaffilmark{3}, Young Sun Lee\altaffilmark{6}, Projjwal Banerjee\altaffilmark{7}, Dmitrii Gudin\altaffilmark{4,5}, Kohei Hattori\altaffilmark{8}, Haining Li\altaffilmark{9}, Tadafumi Matsuno\altaffilmark{10,11}, Vinicius M.  Placco\altaffilmark{4,5},  M. C. Smith\altaffilmark{1}, Devin D. Whitten\altaffilmark{4,5}, Gang Zhao\altaffilmark{9,12}}

\altaffiltext{1}{Key Laboratory for Research in Galaxies and Cosmology, Shanghai Astronomical Observatory, Chinese Academy of Sciences, 80 Nandan Road, Shanghai 200030, China; sala.yuan@gmail.com}
\altaffiltext{2}{Harvard Smithsonian Center for Astrophysics, Cambridge, MA 02138, USA}
\altaffiltext{3}{Institute of Astronomy, University of Cambridge, Madingley Road, Cambridge CB3~0HA, UK}
\altaffiltext{4}{Department of Physics, University of Notre Dame, Notre Dame, IN. 46556, USA}
\altaffiltext{5}{JINA Center for the Evolution of the Elements (JINA-CEE), USA}
\altaffiltext{6}{Department of Astronomy and Space Science, Chungnam National University, Daejeon 34134, Republic of Korea}
\altaffiltext{7}{Department of Physics, Indian Institute of Technology Palakkad, Palakkad, Kerala 678557, India}
\altaffiltext{8}{Department of Physics, Carnegie Mellon University, 5000 Forbes Avenue, Pittsburgh, PA 15213}
\altaffiltext{9}{Key Lab of Optical Astronomy, National Astronomical Observatories, Chinese Academy of Sciences, A20 Datun Road, Chaoyang, Beijing 100102, China}
\altaffiltext{10}{Department of Astronomical Science, School of Physical Sciences, SOKENDAI (The Graduate University for Advanced Studies), Mitaka, Tokyo 181-8588, Japan}
\altaffiltext{11}{National Astronomical Observatory of Japan (NAOJ), 2-21-1 Osawa, Mitaka, Tokyo 181-8588, Japan}
\altaffiltext{12}{School of Astronomy and Space Science, University of Chinese Academy of Sciences, No.19(A) Yuquan Road, Shijingshan District, Beijing, 100049, China}

\begin{abstract}
We search for dynamical substructures in the LAMOST DR3 very metal-poor (VMP) star catalog. After cross-matching with {\it Gaia} DR2, there are $\sim$ 3300 VMP stars with available high-quality astrometric information that have halo-like kinematics. We apply a method based on self-organizing maps, \textsc{StarGO}, to find groups clustered in the 4D space of orbital energy and angular momentum. We identify 57 dynamically tagged groups, which we label DTG-1 to DTG-57. Most of them belong to existing massive substructures in the nearby halo, such as the {\it Gaia} Sausage or Sequoia. The stream identified by Helmi et al. is recovered, but the two disjoint portions of the substructure appear to have distinct dynamical properties. The very retrograde substructure Rg5 found previously by Myeong et al. is also retrieved. We report six new DTGs with highly retrograde orbits, two with very prograde orbits, and 12 with polar orbits. By mapping other datasets (APOGEE halo stars, and catalogs of $r$-process-enhanced and CEMP stars) onto the trained neuron map, we can associate stars with detailed chemical abundances to the DTGs, and look for associations with chemically peculiar stars. The highly eccentric $Gaia$ Sausage groups contain representatives both of debris from the satellite itself (which is $\alpha$-poor) and the Splashed Disk, sent up into eccentric halo orbits from the encounter (and is $\alpha$-rich). The new prograde substructures also appear to be associated with the Splashed Disk. The DTGs belonging to the $Gaia$ Sausage host two relatively metal-rich $r$-II stars and six CEMP stars in different sub-classes, consistent with the idea that the $Gaia$ Sausage progenitor is a massive dwarf galaxy. Rg5 is dynamically associated with two highly $r$-process-enhanced stars with [Fe/H] $\sim -$3. This finding indicates that its progenitor might be an ultra-faint dwarf galaxy that has experienced $r$-process enrichment from neutron star mergers. 
\end{abstract}

\keywords{galaxies: halo --- galaxies: kinematics and dynamics --- galaxies: formation --- methods: data analysis}

\section{Introduction}
\label{intro}

The proto Milky-Way halo has undergone frequent mergers with small dwarf galaxies in the early universe (redshift $z$ $\ga$ 1). Some dwarf galaxies have survived until today, such as the Fornax and Sculptor dwarf spheroidal galaxies (dSphs), 
while some of the survivors are being shredded by the Milky Way and formed tidal streams, such as the Sagittarius dwarf galaxy \citep{ibata95, mateo96, ibata01, maj03}, seen clearly in the \lq Field of Streams' \citep{Be06}. 
Others, especially the most ancient and least massive dwarfs, are fully disrupted and can only be traced from their debris. \citet{helmi99} found a significant substructure of eight stars from Hipparcos data on a highly inclined orbit, clumped in phase space and metallicity; additional candidate members of this stream were added by \citet{chiba00}. \citet{Sm09} used a ``light-motion" catalogue, built from the multi-epoch Sloan Digital Sky Survey (SDSS) Stripe 82 dataset, to find four discrete kinematic over-densities among the halo sub-dwarfs, one of which was the \citet{helmi99} moving group. In addition, an {\it ex situ} or accreted component of the inner halo was suggested by several earlier studies of local halo stars~\citep[e.g.,][]{chiba00,carollo07, carollo10,nissen10,beers12,an15}.

In the {\it Gaia} Era, the magnificent astrometry provided by its Data Releases \citep[DR,][]{Brown18} can be combined with a variety of spectroscopic surveys such as SDSS~\citep{York00}, SEGUE~\citep{yanny09}, RAVE~\citep{Ku17}, APOGEE~\citep{Ab18}, and LAMOST~\citep{cui12,zhao12,deng12} to study the 6D kinematics of millions of stars. Several studies quickly showed that the inner halo is dominated by the single accretion event of a massive progenitor~\citep{belokurov18,myeong18b,haywood18,helmi18}. This structure is known as the {\it Gaia} Sausage \citep[see e.g.,][]{belokurov18}, identified in the SDSS-Gaia DR1 dataset \citep{De17}, or {\it Gaia}-Enceladus \citep{helmi18}, based on Gaia DR2. However, there are differences in the dynamics of the stars included in the two identifications. The {\it Gaia} Sausage comprises stars with high eccentricities and strongly radial motions, with near-zero rotation, consistent with a head-on collision ~\citep{belokurov18}, whereas {\it Gaia}-Enceladus includes many retrograde stars in addition~\citep{helmi18}. 

Judged on dynamical grounds, the retrograde stars partially included as members of {\it Gaia}-Enceladus are more likely to be the residue of a different, but prominent, accretion event, hinted at earlier by ~\citet{carollo07} and \citet{majewski12}. Searches in velocity or action space~\citep{myeong18b,myeong18c} have demonstrated the existence of numerous high-energy, retrograde stellar substructures. More interestingly, many of halo globular clusters on retrograde orbits have specific orbital characteristics (e.g., inclinations, eccentricity) which are also very similar to the retrograde stellar substructures. Based on the numerous retrograde halo substructures with particular orbital characteristics, they are now believed to be associated with a counter-rotating accretion event, called the \lq Sequoia Event' \citep{myeong19}. It has been suggested that Sequoia contributes to the relatively low-eccentricity halo component studied with APOGEE DR14 data \citep{mackereth19}. Using the SAGA database~\citep{Suda08} of metal-poor stars with detailed chemical abundances, \citet{matsuno19} also showed that this high-energy retrograde group has distinct $\mathrm{[\alpha/Fe]}$ abundance patterns compared to the {\it Gaia} Sausage. This adds additional chemical evidence that these retrograde halo stars indeed come from a distinct accretion event, reinforced by the recent analysis of \citet{Mo19}. This interpretation is also supported by the identification of a break radius in the relative age profile of BHB stars in the inner halo, seen to occur at $\sim 14$\,kpc \citep{Whitten2019}, which implies two distinct stellar populations with disparate ages.

In this work, we search for the debris from small, ancient dwarf galaxies. Although they leave detritus in the nearby halo, their contribution is much smaller than massive systems, such as the {\it Gaia} Sausage and Sequoia. Thus, finding their debris is more difficult, because both the number of stars is smaller and the clustering signature is weaker. These obstacles may be overcome by examination of stellar samples in the low-metallicity regime. Small dwarf galaxies are primarily populated by very metal-poor (VMP; [Fe/H] $< -2.0$) stars \citep[e.g.,][]{simon19}, thus the fraction of halo stars originating in low-mass systems is higher in samples of stars having lower metallicities.
    
Here, we examine the largest currently available VMP catalog, from LAMOST DR3 \citep{li18}. Most of the stars in the catalog have heliocentric distances less than 5 kpc, so there are precise parallaxes and proper motions available from {\it Gaia} DR2 \citep{Brown18,gaiadr2}. This allows us to select stars with halo kinematics, and calculate their orbital energies and angular momenta. We look for substructures clustered in dynamical space, using the neural network based clustering method StarGO \citep{yuan18}, which implements an unsupervised learning algorithm, Self-Organizing Maps. Neurons learn the structure from the input data set, and the learning results can be visualized by a 2D map from which groups can be identified and their significance quantified.

The clustering method employed in this work is different from the algorithms used previously, which rely on first building a smooth background model in velocity or action space from the data, and then looking for residuals with respect to the background~\citep{carlin16,myeong18a,myeong18c}. We are interested in comparing the results of our algorithm with earlier catalogs of substructure, in order to provide added confidence in the substructures that are identified, irrespective of the algorithm employed. As VMP stars also are found in larger dwarf galaxies, we also expect to identify massive accreted systems like the {\it Gaia} Sausage~\citep{belokurov18} and Sequoia \citep{myeong19}, based on their characteristics in phase space and orbital properties.

Some of the VMP substructures are the remnants of ultra-faint dwarfs (UFDs) and dwarf spherodials (dSphs). However, they are much closer than any surviving intact satellites, the nearest of which are at least $\sim$ 20 kpc from the Galactic center. The VMP groups therefore offer a unique opportunity to study the chemistry of UFDs and dSphs with nearby, relatively bright, stars. The UFDs provide an environment in which early nucleosynthesis signatures are recorded in 
$r$-process-enhanced stars and different sub-classes of carbon-enhanced metal-poor (CEMP) stars. Currently, the UFD Reticulum II is known to be enriched by rapid neutron-capture events, as 7 out of 9 of its observed giant stars are very $r$-process-enhanced stars \citep{ji16, roederer16}. Recently, the UFD Tucana III has been shown by \citet{Marshall19} to possess at least four additional moderately $r$-process-enhanced stars, in addition to the one originally identified by \citet{hansen17}.  The other UFDs exhibit extreme deficiencies in their heavy-element abundances, lacking over-abundances of both $s$-process and $r$-process elements. Among the canonical dSphs, almost all of them have some moderately enhanced $r$-I (i.e., +0.3 $\leqslant$ [Eu/Fe] $\leqslant +1.0$) and highly enhanced $r$-II (i.e., [Eu/Fe] $>$ +1.0) stars. However, there are only a few stars for each galaxy having their full set of elemental-abundance patterns measured \citep[see][and references therein]{hansen17}. 

Here, we show how to search for possible associations between disrupted satellite debris and chemically peculiar stars. At present, it is challenging to trace their origins due to the small numbers of stars with available high-resolution confirmation of their distinctive nucleosynthetic patterns, although previous attempts have provided interesting results already (e.g., \citealt{roederer18,hansen19}). However, as an alternative approach, we can map these catalogs onto the trained neural network to look for associations between our dynamically tagged groups and chemically peculiar stars~\citep[c.f.,][]{yuan19}. This opens the window to the study of the birth environment of these stars in the nearby halo, where high-resolution spectroscopic data are readily obtainable. The same method can be used to map abundances, such as those for the halo sample of APOGEE~\citep{Ab18}, onto the dynamical groups to study their positions in the plane of abundance versus metallicity, and hence study their relationship to the samples of thin- and thick-disk stars from the same data set.

This paper is organized as follows. Details of our input catalog of VMP stars are described in Sec.~\ref{sec:data}. The methodology of the group identification and the association with stars from other catalogs are presented in Sec.~\ref{sec:method}. We compare the identified dynamically tagged groups with existing substructures in Sec.~\ref{sec:vmpgrp}, emphasising our new discoveries. Chemically peculiar stars associated with our substructures are discussed in Sec.~\ref{sec:chem}. Finally, we summarize our principal results in Sec.~\ref{sec:diss}.
\section{Data}
\label{sec:data}

The LAMOST DR3 VMP catalog is the largest (published) sample of VMP stars to date, containing 10,008 stars covering a large area of sky in the Northern Hemisphere~\citep{li18}. We cross-match LAMOST DR3 VMP with {\it Gaia} DR2, and obtain 9690 stars with full 6-D kinematics. We then employ a pipeline similar to the SEGUE Stellar Parameter Pipeline (SSPP; \citealt{lee08,beers14}), modified to work with LAMOST data \citep{Lee15}, to obtain refined estimates of the atmospheric parameters $T_{\rm eff}$, log g, and [Fe/H] (as well as [C/Fe] and [$\alpha$/Fe], although we do not make use of these abundance ratios in the present paper).  In the process, we have identified and eliminated several thousand stars from the original catalog which were incorrectly identified as VMP for a variety of reasons -- usually poor flux calibration, spectral defects in the region of the Ca~II K line, unusually high reddening estimates, etc.. There are a total of 7814 stars that remain with [Fe/H] $< -1.8$.\footnote{A ``cleaned" LAMOST DR3 VMP catalog is provided in an online dataset accompanying this paper.}

We correct the $Gaia$ parallax measurement for each star by adding an offset of 0.054 mas~\citep{Sc19,Ev19}, and use the inverse of the corrected parallax to calculate the distance. We further select stars within 5 kpc with precise parallax measurements. The radial velocity is corrected by 4.75 kms$^{-1}$ from the cross-match between LAMOST DR3 VMP and {\it Gaia} RVS. The values of solar motion used are ($U$, $V$, $W$) = (11.10, 12.24, 7.25) kms$^{-1}$~\citep{Sc19}, and the motion of the LSR is $v_{\rm LSR}$ = 232.8 kms$^{-1}$~\citep{McM11}. Since we are interested in halo stars from an accreted origin, the Toomre diagram is used to select those with clear halo kinematics. All the selection rules are summarized below:
\begin{enumerate}
\item \texttt{parallax} $\geq$ 0.2 mas,
\item \texttt{parallax\_over\_error} $\geq$ 5,
\item $\|$\textbf{v}-\textbf{v}$_{\rm LSR}\| >$ 210 km/s.
\end{enumerate}
Application of the above criteria yields 3364 stars, which we refer to below as the VMP halo catalog. Using the same criteria, we create a halo sample of 4546 stars from APOGEE data \citep{Ab18}. We first use the VMP halo sample for substructure searching. After candidate groups are identified, we try to find their associated members in the APOGEE halo sample and samples of chemically peculiar stars. Specifically, the latter includes the $r$-process-enhanced star catalog from \citet{roederer18} and the CEMP catalog of stars provided by \citet{yoon16}, with a number of additional supplements from the literature.  A less strict parallax cut (\texttt{parallax\_over\_error} $\geq$ 3) is imposed for pruning these two samples, yielding 79 and 202 stars, respectively.

\begin{figure*}
\includegraphics[width=\linewidth]{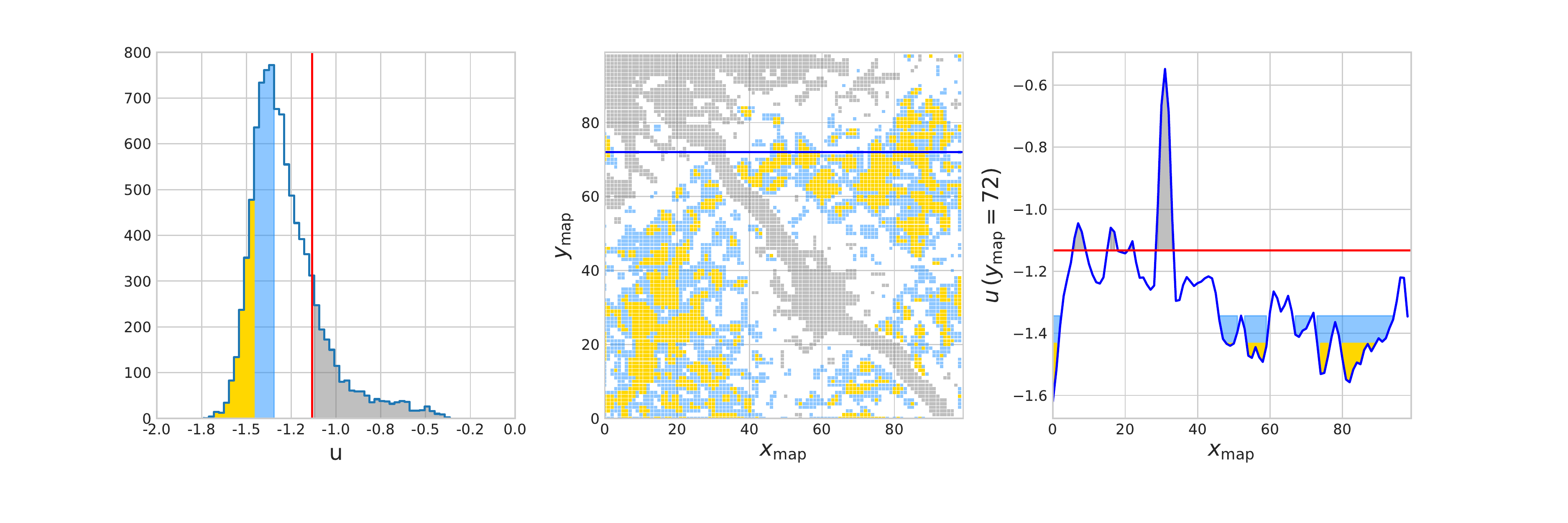}
\caption{Results from the application of \textsc{StarGO} to the VMP halo catalog in the normalized ($E$, $L$, $\theta$, $\phi$) space. Left: The histogram of $u$ values (or separations in weight vectors between adjacent neurons), where the yellow, light blue, and gray areas denote $u\leqslant u_{16\%}$, $u\leqslant u_{40\%}$, and $u\geqslant u_{80\%}$, respectively. Middle: The 100x100 self-organizing map with selected neurons marked with the same color coding. The blue solid line denotes the slice at $y_{\mathrm{map}}$ = 72 as a typical representative. Right: The cross-section $x_{\mathrm{map}}$-$u$ of the slice for $y_{\mathrm{map}}$ = 72, with the red line representing $u$ = $ u_{80\%}$. The areas above the curve and under $u_{16\%}$  are yellow, under $u_{40\%}$ are light blue, while those below the curve and above $u_{80\%}$ are gray. }
\vspace{1cm}%
\label{fig:som}
\end{figure*}
\begin{figure}
\centering
\includegraphics[width=\linewidth]{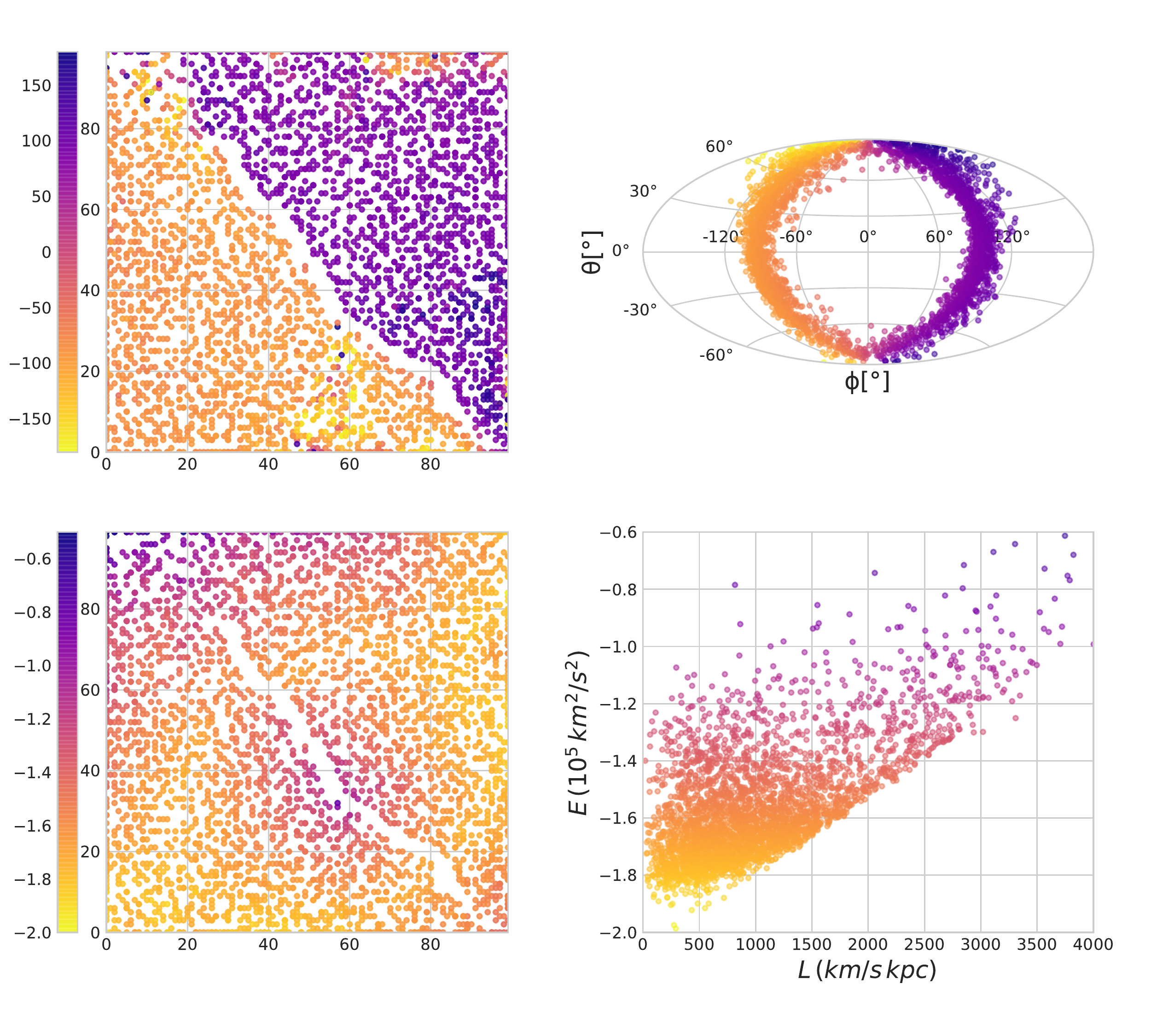}
\caption{The relationship between the map of neurons (left) and the input space (right). In the upper panels, the VMP halo sample is color coded by $\phi$, which describes the phase of the orbit as in eq~(\ref{eq:angles}); in the lower panels, the color coding is by energy $E$. Notice that the nearly diagonal line provides a good separation between orbits with positive and negative $\phi$ in the upper panels. Notice also that the low-energy orbits ($E \lesssim -$1.6 km$^2$s$^{-2}$) are present in the same regions as the light blue patches in Fig.~\ref{fig:som}.}
\label{fig:bnd}
\end{figure}

\section{Method}
\label{sec:method}

\subsection{Prescriptions}
\label{subsec:pres}

We employ the neural network based clustering method \textsc{StarGO} \citep{yuan18} to find substructures that are clustered in energy and integrals of motion space. We calculate the orbital energy in the gravitational potential of \citet{mc17} using \textsc{AGAMA}~\citep{agama}. The energy ($E$) and components of the angular momentum ($L_x, L_y, L_z$) are approximately conserved, even if the potential is not spherical~\citep[e.g.,][]{helmi00}. Specifically, a network of 100$\times$100 neurons are trained by the VMP halo catalog in the ($E$, $L$, $\theta$, $\phi$) space, where 
\begin{equation}
    \theta = \arccos(L_z/L), \qquad\qquad
    \phi = \arctan(L_x/L_y).
    \label{eq:angles}
\end{equation}
We briefly summarize the neural network learning process and group identification procedure below; a more detailed description is available in \citet{yuan19}. 

Each grid point of the map shown in the middle panel of Fig.\ref{fig:som} hosts a neuron, which has a weight vector with the same dimension as the input vector. The values for all the weight vectors are initially randomized. For any given star from the sample, we find the neuron having the closest weight vector to its input vector. This neuron is referred to as its best-matching unit (BMU). The unsupervised learning is performed by updating the weight vector of every neuron of the map closer to the input vector of the given star. The learning effectiveness of each neuron depends on its distance from the BMU on the map. The neighboring neurons of the BMU learn more effectively by updating their weight vectors much closer to the input vector compared to neurons located farther away. All the neurons change their weight vectors from the previous values for every input star. One iteration finishes after all the stars are input into the neural network once. After a sufficient number of iterations, the weight vectors converge and the final map becomes self-organized. Therefore, the entire learning process is called a Self-Organizing Map. The difference in weight vectors between neighboring neurons is defined as $u$. The trained neural network can be quantified by a 100$\times$100 matrix of $u_{\mathrm{mtx}}$. The distribution of all the elements of the matrix is shown in the left panel of Fig.~\ref{fig:som}.

In the middle panel of Fig.~\ref{fig:som}, we highlight neurons with $u<u_{16\%}$ and $u<u_{40\%}$ with yellow and light blue, respectively. They represent neurons which, judged by similarity, lie in the top 16$\%$ and 40$\%$, respectively. For comparison, neurons with  $u>u_{80\%}$ are colored gray; these have similarities in weight vectors that lie in the bottom 20$\%$. The different sets of neurons are associated with stars having correspondingly similar or dissimilar features in the input space. The most prominent structure revealed by the trained neuron map is the nearly diagonal boundary. Most of the neurons with high similarities (the yellow and light blue patches) reside in the lower left and upper right corners separated by the gray boundary line. Neurons in these two regions correspond to stars with negative and positive $\phi$, which is the azimuthal phase angle of angular momentum (see the first row of Fig.~\ref{fig:bnd}). Stars in the region of low absolute values of $\theta$ (inclination angle of the angular momentum) are clearly separated into two sets, depending on their values of $\phi$. The angular momentum of these stars are in nearly opposite directions, thus their BMUs sit on opposite sides of the nearly diagonal boundary line. Those stars with high absolute values of $\theta$ are close in angular momentum directions; they are found in the upper region of the neuron map in the upper-left panel of Fig.~\ref{fig:bnd}.

Besides the diagonal boundary line, we can also see that the boundaries separate several white patches into isolated islands (see the upper left corner in the middle panel of Fig.~\ref{fig:som}). Neurons in the islands have more similar weight vectors compared to those in their surroundings. We take a slice of the neuron map at $y_{\mathrm{map}}=72$, shown as the blue line, and plot the values of $u$ along the grid point of $x_{\mathrm{map}}$ in the right panel of Fig.~\ref{fig:som}. We fill the area above the curve with $u<u_{16\%}$ and $u<u_{40\%}$ with yellow and light blue, respectively. These regions represent the range of $x_{\mathrm{map}}$ of the slice crossing the colored patches of the neuron map. Similarly, the area under the curve with $u>u_{80\%}$ is filled with gray, denoting the range of the slice in the gray region. The red horizontal line shows the defined boundary at $u_{80\%}$. As can be seen, the isolated islands containing both light blue and yellow patches around the left edge of the map ($x_{\mathrm{map}}$ = 0, $y_{\mathrm{map}}$ $\sim$ 72) correspond to the deep valley in the ($x_{\mathrm{map}}$, $u$) plot. 

The group identification algorithm is performed by gradually decreasing the value of the defined boundary until isolated islands appear in the middle panel of Fig.~\ref{fig:som}. This is equivalent to lowering the red line in the right panel until the valley is separated by gray peaks for slices from all directions across that island. Intuitively, we consider the input data space is composed of star clusters overlapping with each other. By taking a 1D slice of the neuron map, they are shown as valleys. If the valleys have lower value of $u$ in slices from all directions, they correspond to isolated islands in the neuron map. The goal of the group identification is to find the critical value of $u$ which can pick out these structures. We then try to obtain the most complete member list for each group, and check its significance and level of contamination for validation during the second step (see details in Sec.~\ref{subsec:map}). We begin to find islands at the 90th percentile of the distribution of $u$, $u_{90\%}$, and perform the group identification all the way to the 5th percentile, $u_{5\%}$.

\begin{figure*}[ht]
\centering
\includegraphics[width=\linewidth]{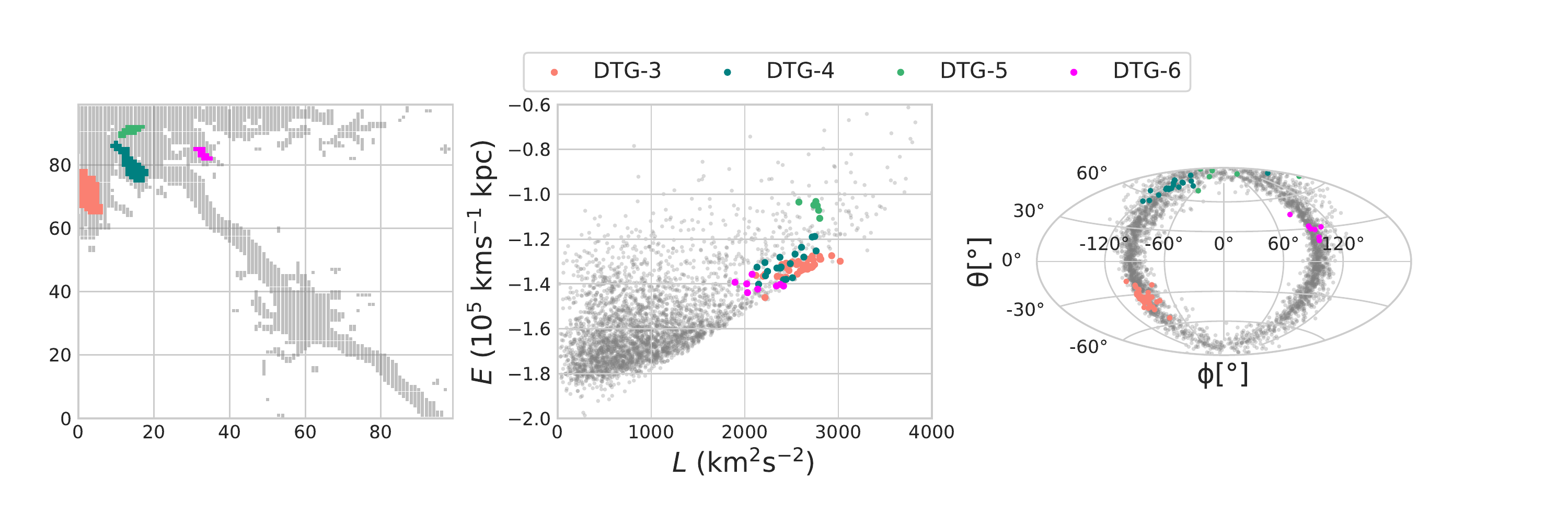}
\caption{Example of the group-identification process. Here, we use the threshold value $u_{\mathrm{thr}}$ = $u_3 = 80 \%$, which is the same as shown in Fig.~\ref{fig:som}. Four groups are identified in the neuron map in the left panel, namely DTG-3 (salmon), DTG-4 (dark green), DTG-5 (light green) and DTG-6 (magenta).  The middle and right panels show the input space of energy $E$ and total angular momentum $L$, as well as direction of the pole of the angular momentum vector. The four groups are clustered in dynamical space. Though there is some overlap in the projection in ($E,L$), the groups are well-separated in the direction of their angular momentum vectors ($\theta,\phi$). The nearby halo stars have $\phi$ in the range of $-$60$^{\circ}$ to $-$120$^{\circ}$ and 60$^{\circ}$ to 120$^{\circ}$, because the position vectors of these stars are close to $\phi$ = 0$^{\circ}$.}
\label{fig:gi}
\end{figure*}

\subsection{Mapping Different Data Sets}

\label{subsec:map}

After the unsupervised learning process, we can map different datasets to the trained neural network by matching input vectors to the BMUs. By this means, we are able to obtain the relationship between the new dataset and the trained network. The methodology of this procedure is illustrated in detail in \citet{yuan19}, where it is used to show that the Cetus Stream is associated with the globular cluster NGC 5824. Here, we apply the same approach to map additional catalogs (APOGEE, $r$-process-enhanced stars, and CEMP stars) onto the trained map, as well as to quantify the significance, contamination, and confidence level of the identified groups. 

The algorithm is summarized in the following steps. For a given group $\mathcal{G}$, which has $n \geqslant 5$ star members, extracted from the VMP sample ($\mathcal{S}$) with a total $\mathcal{N}$ stars, we proceed as follows:

\begin{enumerate}
\item Obtain the probability density functions (PDFs) for energy $E$, modulus of angular momentum $L$ and its direction $\theta$ and $\phi$ by Gaussian kernel density estimation.

\item Draw a random sample ($\mathcal{S}$1) of 5000 stars according to the PDFs from 1.

\item Find the BMU for every star from $\mathcal{S}$1 on the trained neuron map, and obtain the stars associated with group $\mathcal{G}$ on the map.

\item Calculate the 4-D distance between member stars of $\mathcal{G}$ and their BMUs. Obtain the largest distance for $\mathcal{G}$ denoted by $u_{\mathrm{vw,max}}$.

\item Calculate the 4-D distance between stars from $\mathcal{S}$1 associated with $\mathcal{G}$ and their BMUs, which is referred to as $u_{\mathrm{vw}}$. Retain the stars with $u_{\mathrm{vw}}\leqslant u_{\mathrm{vw,max}}$, $n_1$. Thus the probability of the retained stars is $p = n_1/5000$.

\item Calculate the binomial probability $\mathcal{P}$ of detecting a group with more than $n$ stars from the total sample of $\mathcal{N}$ stars, given the probability $p$. If 1 - $\mathcal{P}$ $\geqslant$ 99.73$\%$, the significance of $\mathcal{G}$ is larger than 3$\sigma$, and we consider it as a possible detected group.

\item If $\mathcal{G}$ is a possible detected group, we estimate the contamination fraction from $\mathcal{S}$1, which is defined as $\mathcal{F}_{c} = p/(n/\mathcal{N})$. $\mathcal{G}$ is considered as valid if $\mathcal{F}_{c}\leqslant$ 40$\%$.
\end{enumerate}


After obtaining all of the valid groups, the same approach is utilized to check the confidence of membership. For $\mathcal{G}$, we first generate 100 Monte Carlo realizations for each group member and denote this sample as $\mathcal{S}_{\mathrm{MC}}$. This is done by considering the observational uncertainties in the 5D astrometric parameters together with the covariance matrix, as well as the uncertainty in radial velocity. Applying steps 1 -- 5 to $\mathcal{S}_{\mathrm{MC}}$, we obtain the probability ($p_{\mathrm{MC}}$) of the Monte Carlo realizations that are associated with $\mathcal{G}$, which quantifies the confidence level of a given group member. 
The overall confidence level of $\mathcal{G}$ is estimated by averaging $p_{\mathrm{MC}}$ for all the members, as listed in Table~\ref{tab:vmpgrp}. For example, DTG-52 has the lowest confidence level, 39$\%$, among all the groups. This value means that its groups members have a 39$\%$ chance, on average, to be identified as members of DTG-52, taking into account the observational errors.


Adopting the same approach, the trained neuron map enables us to find the associated group members from the APOGEE halo sample, as well as from the $r$-process-enhanced star and CEMP star catalogs. Similarly, we generate 1000 Monte Carlo realizations for each star in a new sample, and apply steps 1 -- 5 to that. We obtain the probability ($p_{\mathrm{APOGEE}}$, $p_{\mathrm{r}}$, and $p_{\mathrm{CEMP}}$) of the Monte Carlo realizations that are associated with $\mathcal{G}$. We consider the star to be a valid member of $\mathcal{G}$ if its probability is greater than 25$\%$.

\begin{table*}
\centering
\caption{VMP Groups}\label{tab:vmpgrp}
\def\arraystretch{1.5}
\begin{minipage}{3.5in}

\begin{tabular}{|c|ccccc|}
	\hline
	 $u_{\rm{thr}}$ &     $\mathcal{G}$     &  $n$    &     $\mathcal{F}_{\mathrm{c}}$  &  Confidence  & Substructure  \\
\hline

 $u_{1}$(90$^{\rm th}$)

& DTG-1 &  9 & 0 \% & 85 \% & Helmi ? ($v_z>0$) \\
\hline

 $u_{2}$(84$^{\rm th}$)  

& DTG-2 &  8 & 0 \% & 82 \% & Cand14 ? \\
\hline

 $u_{3}$(80$^{\rm th}$) 

& DTG-3 &  38 & 1 \% & 94 \% & Helmi ($v_z<0$) \\
& DTG-4 &  20 & 37 \% & 98 \% & Sequoia \\
& DTG-5 &  6 & 0 \% & 87 \% & Sequoia \\
& DTG-6 &  8 & 8 \% & 92 \% & Rg5 \\
\hline

 $u_{4}$(70$^{\rm th}$)

& DTG-7 &  8 & 25 \% & 69 \% & Sausage \\
& DTG-8 &  7 & 19 \% & 78 \% & New, Polar \\

\hline

 $u_{5}$(65$^{\rm th}$)

& DTG-9 &  9 & 33 \% & 68 \% & Sausage \\
\hline

 $u_{6}$(57$^{\rm th}$)

& DTG-10 &  5 & 0 \% & 70 \% & Rg5 \\
& DTG-11 &  15 & 0 \% & 91 \% & Cand10 ? \\
& DTG-12 &  7 & 19 \% & 91 \% & Sausage \\
& DTG-13 &  15 & 22 \% & 99 \% & Sausage \\
& DTG-14 &  10 & 13 \% & 81 \% & Sausage \\
& DTG-15 &  7 & 9 \% & 78 \% & New, Polar \\
\hline

 $u_{7}$(50$^{\rm th}$)

& DTG-16 &  5 & 0 \% & 80 \% & Sausage \\
& DTG-17 &  7 & 8 \% & 58 \% & Sausage \\
& DTG-18 &  16 & 16 \% & 66 \% & Sausage \\
\hline

 $u_{8}$(45$^{\rm th}$) 

& DTG-19 &  25 & 28 \% & 86 \% & New, Pg \\
\hline

 $u_{9}$(40$^{\rm th}$) 
& DTG-20 &  4 & 13 \% & 66 \% & Sausage \\
& DTG-21 &  44 & 33 \% & 93 \% & New, Rg \\
& DTG-22 &  27 & 29 \% & 94 \% & New, Rg \\
& DTG-23 &  7 & 19 \% & 79 \% & Rg5 \\
& DTG-24 &  6 & 11 \% & 77 \% & New, Rg \\
\hline
$u_{10}$(30$^{\rm th}$)  
& DTG-25 &  6 & 22 \% & 71 \% & Sausage \\
& DTG-26 &  9 & 22 \% & 90 \% & Sausage \\
& DTG-27 &  13 & 31 \% & 79 \% & Sausage \\
& DTG-28 &  20 & 16 \% & 95 \% & New, Rg \\

\hline
\end{tabular}
\end{minipage}
\begin{minipage}{3.5in}
\begin{tabular}{|c|ccccc|}
	\hline
	 $u_{\rm{thr}}$ &     $\mathcal{G}$     &  $n$    &     $\mathcal{F}_{\mathrm{c}}$  &  Confidence & Substructure  \\
    \hline
$u_{10}$(30$^{\rm th}$)  
& DTG-29 &  74 & 35 \% & 93 \% & New, Rg \\
& DTG-30 &  15 & 26 \% & 86 \% & Sausage \\
& DTG-31 &  25 & 24 \% & 95 \% & Sausage \\
& DTG-32 &  14 & 14 \% & 81 \% & Sausage \\
& DTG-33 &  37 & 25 \% & 88 \% & New, Rg \\

\hline
 $u_{11}$(20$^{\rm th}$)  

& DTG-34 &  8 & 0 \% & 83 \% & New, Polar \\
& DTG-35 &  11 & 6 \% & 85 \% & New, Polar \\
& DTG-36 &  8 & 8 \% & 82 \% & New, Polar \\
& DTG-37 &  5 & 13 \% & 65 \% & Sausage \\
& DTG-38 &  171 & 31 \% & 93 \% & Sausage \\
& DTG-39 &  8 & 22 \% & 97 \% & New, Polar \\
& DTG-40 &  17 & 18 \% & 82 \% & Sausage \\
& DTG-41 &  22 & 9 \% & 94 \% & Sausage \\
& DTG-42 &  11 & 12 \% & 84 \% & Sausage \\

\hline

 $u_{12}$(10$^{\rm th}$)  

& DTG-43 &  8 & 14 \% & 84 \% & New, Polar \\
& DTG-44 &  14 & 9 \% & 87 \% & New, Polar \\
& DTG-45 &  25 & 15 \% & 85 \% & New, Polar \\
& DTG-46 &  25 & 10 \% & 90 \% & Sausage \\
& DTG-47 &  9 & 26 \% & 87 \% & Sausage \\
& DTG-48 &  14 & 17 \% & 88 \% & Sausage \\
& DTG-49 &  5 & 0 \% & 91 \% & Sausage \\
& DTG-50 &  18 & 11 \% & 74 \% & Sausage \\
& DTG-51 &  7 & 8 \% & 68 \% & New, Polar \\
& DTG-52 &  4 & 0 \% & 39 \% & Rg5 \\
& DTG-53 &  5 & 11 \% & 56 \% & Rg5 \\
& DTG-54 &  4 & 0 \% & 97 \% & New, Polar \\
& DTG-55 &  9 & 22 \% & 82 \% & Sausage \\
& DTG-56 &  5 & 0 \% & 89 \% & Sausage \\
\hline

 $u_{13}$(5$^{\rm th}$)  

& DTG-57 &  17 & 3 \% & 80 \% & Sausage \\

	\hline
\end{tabular}
\end{minipage}

  \medskip
Notes -- The values of $u_{\rm thr}$ denoted by the percentiles in the brackets are listed in the first column. For each group $\mathcal{G}$, $n$ is the number of valid members, $\mathcal{F}_{\rm c}$ denotes the contamination fraction, and the confidence level is defined in Section~\ref{subsec:map}. New groups are classified into retrograde (Rg), polar, and prograde (Pg). Two groups may be associated with candidate groups found by \citet{myeong18a}.

\end{table*}

\section{Groups in the VMP catalog}
\label{sec:vmpgrp}

We identify 57 valid groups, at thirteen different values of $u_{\rm{thr}}$, which satisfy the requirements in steps 6 and 7 above. We summarize the properties for all of the groups in Table~\ref{tab:vmpgrp}. Since they are identified dynamically, we refer to them as ``Dynamically Tagged Groups'' (DTGs)\footnote{In order to preserve a meaningful nomenclature in the future, as the numbers of DTGs from various analyses grow, we suggest that, within a given paper, authors use DTG-x (where x is an integer from 1 to n). A given group can be identified for later reference by XXYY:DTG-x, where XX are the first and last initial of the first author, and YY are the last two digits of the year of publication (a, b, c, can be added if more than one paper is published with the same lead author in a given year).  For example, DTG-1 in the present paper should be referred to as ZY20:DTG-1.} In Table~\ref{tab:vmpgrp}, we also list the group size $n$, the contamination fraction $\mathcal{F}_{\rm{c}}$, and the confidence level.

As an illustration of the group identification procedure, Fig.~\ref{fig:gi} shows the DTGs identified with $u_3$ = $u_{80\%}$. The gray boundaries in the left panel are defined by neurons with $u\geqslant u_3$, which is the same threshold $u_{80\%}$ as in Fig.~\ref{fig:som}. We only pick out the newly formed islands at $u_3$, and find four of them are valid groups (DTGs-3, 4, 5, and 6). We note that there are several other white isolated islands in the upper region of the neuron map. These have either been identified with previous threshold values, or they do not meet the criteria for valid groups. We show that DTGs-3, 4, 5 and 6 form separate, but partly overlapping, clusters in energy and angular momentum space in the middle panel. However, the separation between the groups is particularly clean, based on the poles of the angular momentum vector, as shown in the right panel. 

\begin{figure*}
\centering
\includegraphics[width=\linewidth]{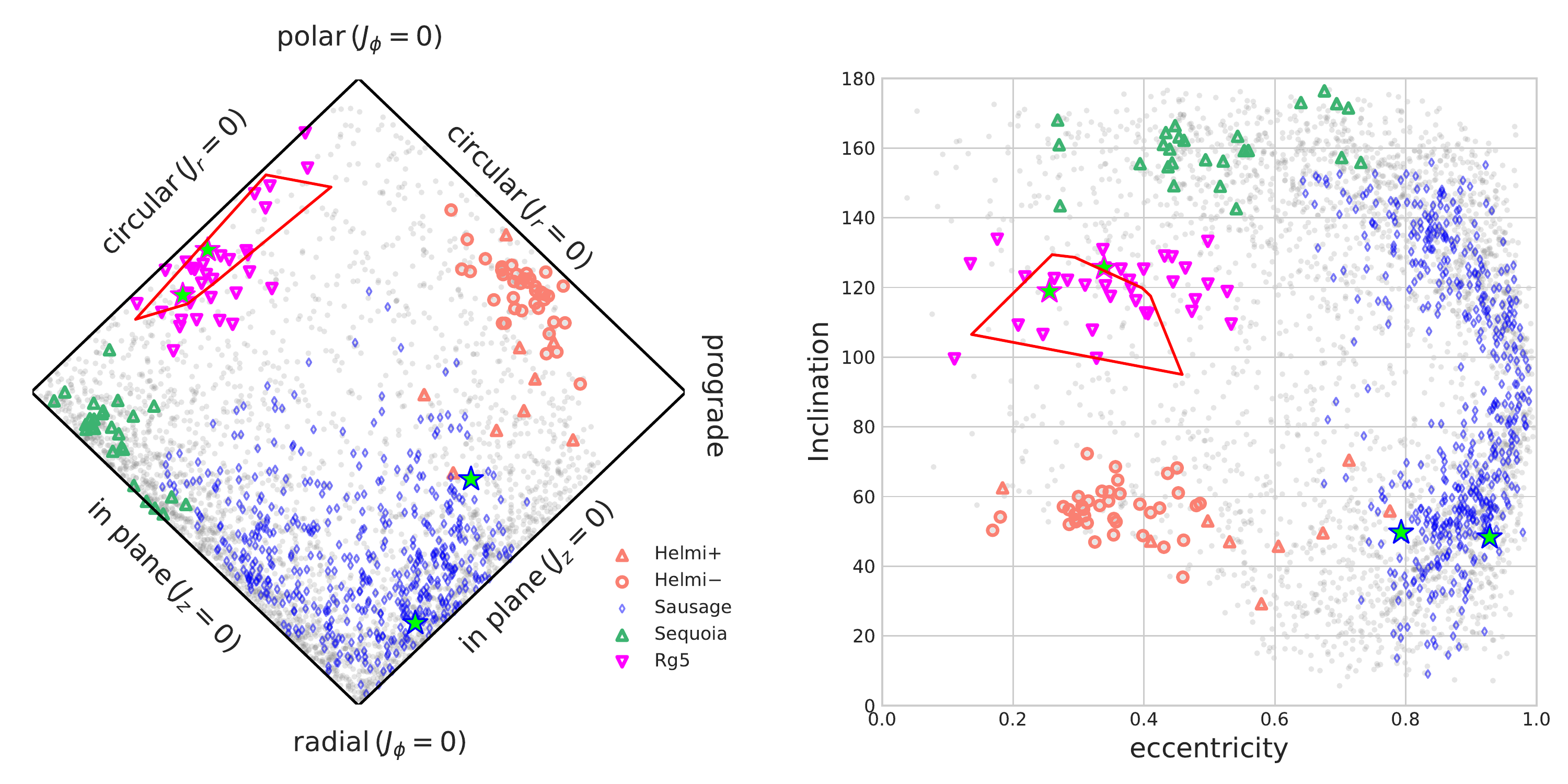}
\caption{The panels show the known VMP substructures in projected action space, and in the space of orbit-averaged eccentricity and inclination angle, as inferred from numerical integration. In the left panel, the $x$-axis is ($J_{\phi}$/$J_{\rm{tot}}$), and the $y$-axis is ($J_{\rm z} - J_{\rm R}$)/$J_{\rm{tot}}$), where $J_{\rm{tot}}$ = $J_{\rm z} + J_{\rm R} + |J_{\phi}|$. The gray dots represent all of the VMP halo stars in the catalog. Note that the stars with disk dynamics are excluded, thus the corner of very prograde orbits in the action space panel is empty. Similarly, the representative region of low $e$ and small $i$ in the space of eccentricity and inclination angle is not occupied. Notice that the DTGs associated with the $Gaia$ Sausage are highly eccentric, while the Sequoia DTGs all have high inclination, consistent with its very retrograde origin. The red polygon boxes show the Rg5 group from \citet{myeong18c}. The two $r$-II stars associated with Rg5 are shown as magenta star symbols with green shading, and are well within the red boxes. The two $r$-II stars associated with the $Gaia$ Sausage are shown as blue star symbols with green shading. The stream discovered by \citet{helmi99} is shown as salmon circles and triangles, depending on the sign of the vertical velocity.}
\label{fig:as_ecc_old}
\end{figure*}

\begin{figure*}
\centering
\includegraphics[width=\linewidth]{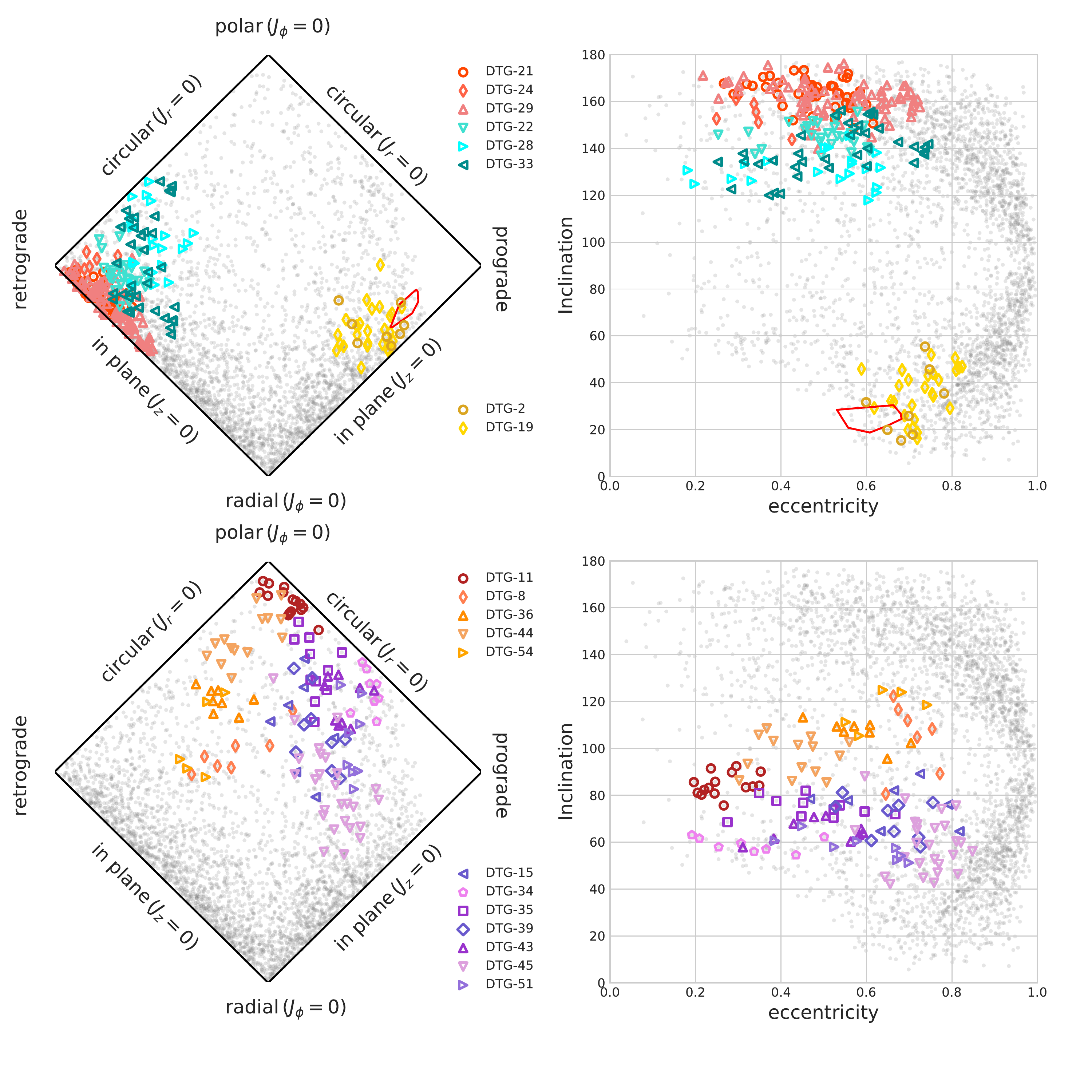}
\caption{The panels show the new VMP groups in the same spaces as in Fig~\ref{fig:as_ecc_old}. These new DTGs separated into predominantly prograde and  retrograde (top panels) and primarily polar (bottom) substructures. The red polygon boxes in the upper panels show Candidate Group 14 from \citet{myeong18c}, which is close to DTG-2.}
\label{fig:as_ecc_new}
\end{figure*}

After the identification, we plot all the groups in the projected action-space map~\citep{vasiliev18}, and the planes of orbit-averaged eccentricity and inclination angle in Figs.~\ref{fig:as_ecc_old} and \ref{fig:as_ecc_new}. The latter quantities are obtained by integrating orbits for about 10 orbital times. The VMP halo stars are shown as gray dots in all the figures. Note that the stars with disk dynamics are excluded, thus the corner of very prograde orbits in the action space is empty. Similarly, the region of stars with low $e$ and small $i$ in the space of eccentricity and inclination angle is not occupied. With these plots in hand, we are able to compare with previous works~\citep{helmi99,belokurov18, myeong18c, myeong19}, and assign origins for the 57 groups. These are named in the last column of Table~\ref{tab:vmpgrp}, and are discussed in detail below. Note that a large substructure like the $Gaia$ Sausage or Sequoia may become fragmented into several smaller groups due to subtle differences in the dynamics -- for example, material torn off at different pericentric passages may have different eccentricities or inclinations because of the effects of dynamical friction. This phenomenon can be also seen in the action plots of earlier works \citep{myeong18c, myeong19}. For example, both DTG-4 and DTG-5 belong to the Sequoia, while numerous groups are all part of the extensive $Gaia$ Sausage debris. The DTGs belonging to the existing substructures are summarized in Table~\ref{tab:sub}. Note that $n_{\rm sub}$ in the first column denotes the total number of members for each substructure. New groups are classified into retrograde (Rg), polar, and prograde (Pg), as shown in Table~\ref{tab:new}.

\subsection{The {\it Gaia} Sausage}

The progenitor of the {\it Gaia} Sausage is believed to be very substantial, possibly comparable in mass to the Large Magellanic Cloud~\citep{belokurov18}. For example, there are up to $\sim 20$ potential globular clusters showing typical chemo-dynamical characteristics of the $Gaia$ Sausage~\citep{myeong18b,myeong19}. The detritus of the $Gaia$ Sausage extends out to Galactocentric radii of $\sim 30$\,kpc \citep{Io19}, which corresponds to the break radius in the stellar halo. \citet{De13} predicted that the break radius was the apocentric pile-up caused by the early accretion of a massive satellite, and this was subsequently confirmed to be the {\it Gaia} Sausage in \citet{De18}. Evidence from the break radius in the relative age profile, as determined from BHB photometry \citep{Whitten2019}, also supports the idea that the inner halo is formed through a few prominent mergers, particularly the $Gaia$ Sausage event. The progenitor galaxy has been wholly destroyed, yet is still only partially phase-mixed in the inner halo, with clouds, shells, and feathers associated with its dismemberment. With such an enormous progenitor, we expect our algorithm to identify substructures within the {\it Gaia} Sausage, as well as detritus flung out of the nascent Milky Way disk by its impact.  

\begin{figure*}[ht]
\centering
\includegraphics[width=\linewidth]{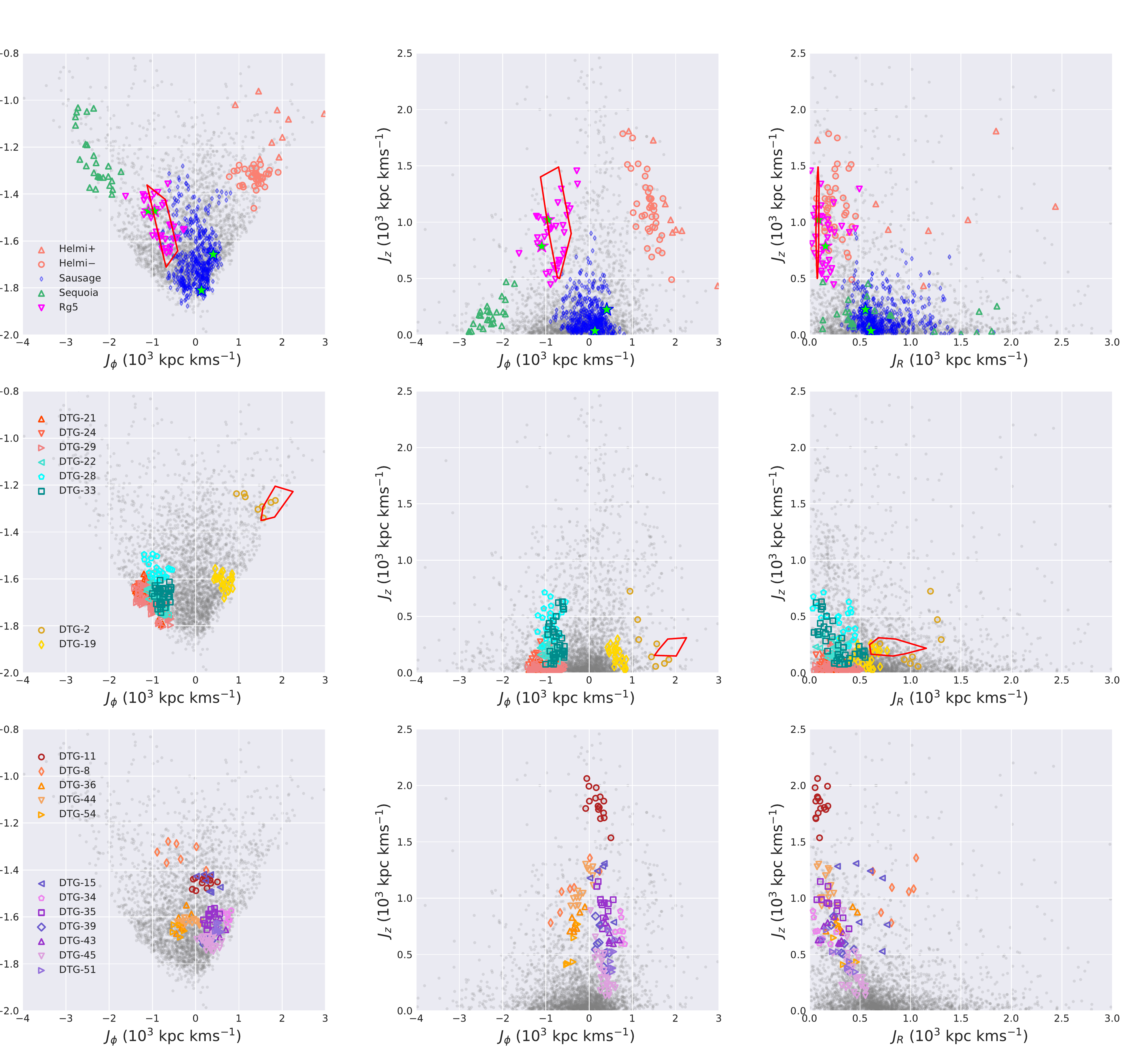}
\caption{Kinematic and dynamic properties of the DTGs, shown in the principal planes of energy and action space. For clarity, the same three plots are repeated in each row, but for different DTGs, as noted in the legends. Notice that the $Gaia$ Sausage substructures (blue diamonds) have zero net azimuthal action, consistent with its interpretation as an almost head-on merger. The $Gaia$ Sausage and Sequoia correspond to the locations seen in \citet{myeong18b} and \citet{myeong19}. The stream found in \citet{helmi99} is separated into two clumps, depending on the sign of the vertical velocity. The red boxes in the panels in the first row show Rg5, while those in the second row show Candidate Group 14 from \citet{myeong18c}. The four $r$-II associated with these substructures are plotted in the same fashion as Fig.~\ref{fig:as_ecc_old}; the two belonging to Rg5 are well within the red box.}
\label{fig:action}
\end{figure*}

At ten different threshold values of $u_{\rm{thr}}$, we find 27 DTGs of 490 stars (see Table~\ref{tab:sub}), all of which have high eccentricities ($\langle e \rangle\ga$ 0.7) and little or no net rotational velocity. We associate DTGs with the $Gaia$ Sausage if they additionally have large radial motions  with $\langle J_{\rm R} \rangle > $ 500 kpc km s$^{-1}$ and small $\langle J_{\rm z} \rangle <$ 500 kpc km s$^{-1}$. This yields the blue groups shown in Figs.~\ref{fig:as_ecc_old} and \ref{fig:action}. Although some other DTGs may have high-eccentricity members with $e\approx 0.7$, we are able to distinguish them from the $Gaia$ Sausage groups based on the action constraints. There are comparable numbers of DTGs that have slightly prograde and retrograde orbits, consistent with the almost head-on encounter originally postulated in \citet{belokurov18}. These DTGs also have different orbital energies, from $-2 \times$10$^5$ to $-1.3 \times$10$^5$ km$^2$s$^{-2}$, consistent with different pieces of the debris from a single, almost radial merger event.

\begin{figure*}
\centering
\includegraphics[width=\linewidth]{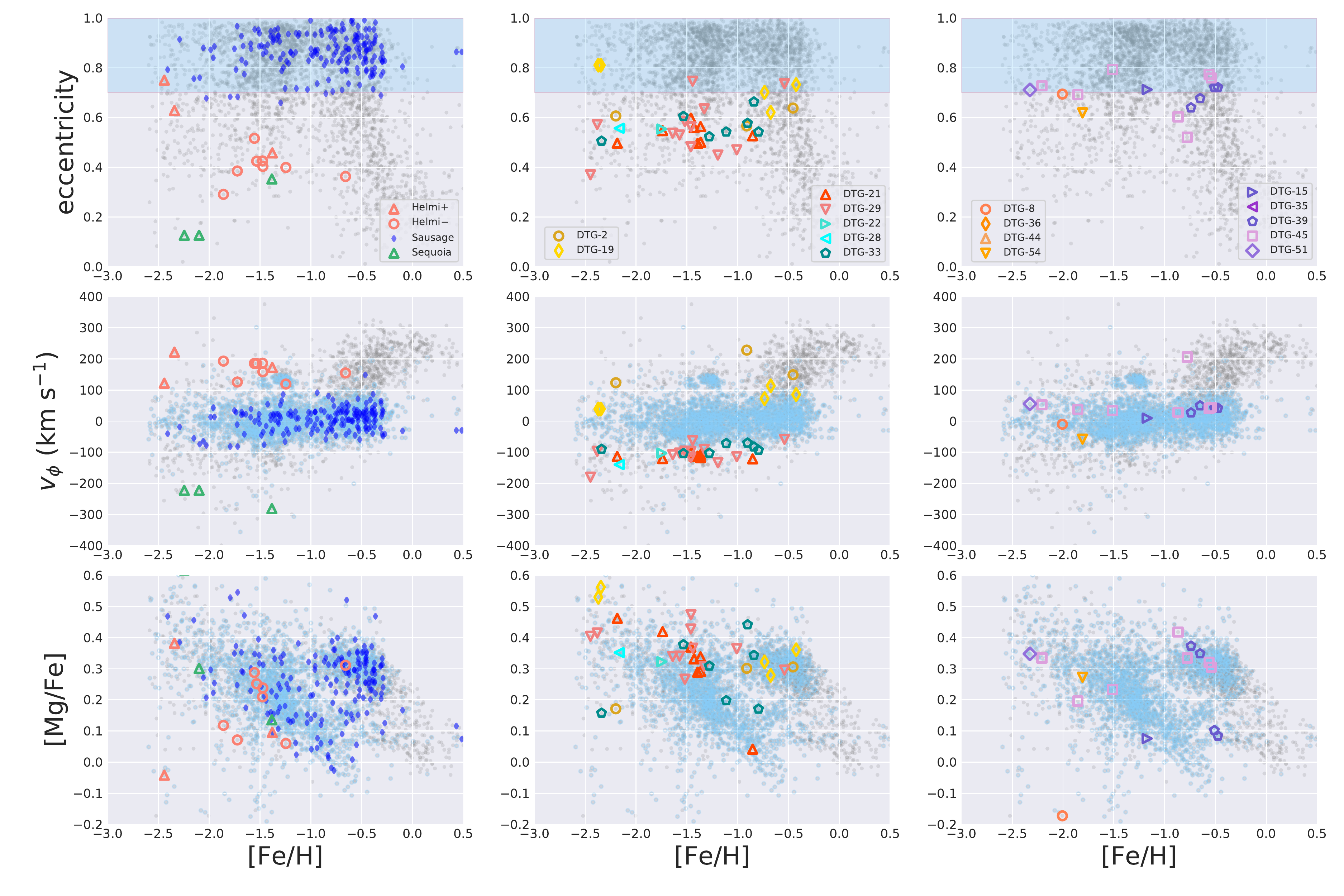}
\caption{Chemo-dynamics of the substructures. Each column shows the planes of eccentricity, azimuthal velocity, and abundance plotted against metallicity, but for different DTGs, as noted in the legends. The gray dots are the APOGEE halo stars, while the blue shaded area corresponds to eccentricity $> 0.7$. These stars are color-coded light blue in the lower two panels. Notice that the high eccentricity stars follow two distinct sequences in the abundance-metallicity plane. All the members associated with DTGs from APOGEE are plotted using the same symbols as before. The small blue diamonds or $Gaia$ Sausage stars are separated into two subgroups, with the more metal-poor members ([Fe/H] $\la -1.3$) showing chemistry consistent with a dwarf galaxy origin, but the more metal-rich members ([Fe/H] $\ga -1.3$) consistent with a thick-disk system origin. These may be stars heated by the $Gaia$ Sausage impact into halo-like orbits, the so-called Splash or Plume~\citep{DiM18,Bel19}. Notice as well that most of the stars belonging to the two prograde groups (DTG-2 and DTG-19) are also consistent with this heated or splashed thick-disk component. The majority of stars belonging to the new retrograde DTGs are metal poor, and follow the low-$\alpha$ sequence of typical dwarf galaxies.}
\label{fig:vphi_ecc_alpha}
\end{figure*}

We check the members of the $Gaia$ Sausage groups from APOGEE data, shown as the blue diamonds in Fig.~\ref{fig:vphi_ecc_alpha}. They clearly separate into two groupings in the metallicity-abundance plot. The subgroups on the low-$\alpha$ track are from the progenitor of the $Gaia$ Sausage. Their chemistry is consistent with that of typical dwarf galaxies. Another subgroup is clustered in the high-$\alpha$ region, which represents the typical chemistry of the thick disk. These stars with high eccentricities and high-$\alpha$ are probably disk stars splashed up by the $Gaia$ Sausage event, evidence for which has been presented elsewhere~\citep{DiM18,Bel19}. Moreover, we highlight the high-eccentricity region ($e\geqslant$ 0.7) by light blue shading in the upper panel, and plot them with the same color in the lower panel. Just as for the $Gaia$ Sausage groups, these high-eccentricity stars are populated in two sequences in the metallicity-abundance plot. For the low-$\alpha$ sequence, there appears a hint of a change in gradient at [Fe/H] = $-1.3$, as also seen by \citet{mackereth19}. High-eccentricity halo stars with [Fe/H] $\ga -1.0$ belong almost wholly to the high-$\alpha$ sequence.


\subsection{Retrograde Groups}

The profusion of retrograde stars in the stellar halo has long been a puzzle~\citep{carollo07, majewski12}. \citet{myeong18c} identified a series of high-significance retrograde substructures in the {\it Gaia} data, which they called Rg1 to Rg7. Subsequently, some of these (Rgs1-4 and 6) were associated with six retrograde globular clusters, including two enormous ones, FSR1758 and $\omega$Centauri. The latter has been suggested by multiple authors to be the core of a nucleated dwarf galaxy~\citep{bekki03,Bek06,Joo13}. This is consistent with the wreckage of another dwarf galaxy -- the Sequoia -- which came in on a strongly retrograde orbit, disgorging globular clusters and debris throughout the inner Galaxy~\citep{myeong19}. 

We find 13 very retrograde groups (DTGs-4, 5, 6, 10, 21, 22, 23, 24, 28, 29, 33, 52, and 53), identified with six threshold values ($u_3$, $u_6$, $u_9$, $u_{10}$, $u_{11}$, and $u_{12}$). By comparing the reported substructures from \citet{myeong19} with the action-space map shown in Fig.~\ref{fig:as_ecc_old}, we are able to clearly associate DTG-4 and 5  with Sequoia (plotted in green symbols), which has 27 stars in total. The Sequoia groups have eccentricity $e \approx 0.5$ and inclination angle $i \approx 160^\circ$. They are also highly retrograde, with $\langle v_\phi \rangle \approx -270$ kms$^{-1}$.

There is another slightly less-retrograde substructure, Rg5, from \citet{myeong19}, shown as red polygonal boxes in Fig.~\ref{fig:as_ecc_old}, as well as in the top row of Fig.~\ref{fig:action}. We find five groups (DTGs-6, 10, 23, 52, and 53) of 33 stars, marked by magenta triangles, that have a very good match with these boxes, and thus are likely associated with Rg5. In contrast to Sequoia, Rg5 has lower eccentricity ($e \approx 0.3$) and inclination angle ($i \approx 120^\circ$), and is less retrograde, with $\langle v_\phi \rangle \approx -90$ kms$^{-1}$. Rg5 has two components of $v_z$, depending on the sign, which are centered at 168 and $-$183 kms$^{-1}$, respectively. Notice too that the Sequoia groups have higher orbital energy than Rg5, so the two substructures are well-separated in the projected space of energy and azimuthal action.

The remaining six retrograde DTGs all have lower orbital energy than Sequoia ($E \leqslant$ -1.6$\times$10$^6$ km$^2$s$^{-2}$), but more retrograde orbits than Rg5, as shown in the ($J_{\phi}$, $E$) space in the second row of Fig.~\ref{fig:action}. Some of these groups may be from the same progenitor, but without chemical information it is hard to be certain. These groups are separated crudely into two sets, colored with coral and cyan, due to their differences in inclination angles, as shown in the upper-right panel of Fig.~\ref{fig:as_ecc_new}. The coral groups (DTGs-21, 24 and 29) have the same mean inclination angle ($i \approx 160^\circ$) as the Sequoia groups. The cyan groups (DTGs-22, 28, and 33) appear to be more diffuse, with slightly lower inclination angles ($i \approx 130^\circ$ -- $145^\circ$). The average rotational velocities of these two sets are $-$120 kms$^{-1}$ and $-$100 kms$^{-1}$, respectively. The coral groups have almost zero mean vertical velocity, whereas the cyan groups have two disjoint portions, with $v_z$ centered at 96 and $-$111 kms$^{-1}$, respectively. Both of these two sets of groups occupy the same corner region of very retrograde orbits as Sequoia in the action-space map (see the right panel of Fig.~\ref{fig:as_ecc_old}). It is possible that some of these DTGs are the low-energy debris from Sequoia.

We also look for the associated APOGEE members for all of the retrograde DTGs. None of the stars can be associated with Rg5, which implies that it is likely a VMP-dominated substructure \citep[e.g.,]{myeong18c}. There are three stars (green triangles) belonging to Sequoia. Two of them have [Mg/Fe] measurements, and sit in the low-$\alpha$ region. We find 20 members for the coral groups, and 9 members for the cyan groups. There are two 
$\alpha$-rich members for both sets, which are stars originated from the disk. Excepting these stars, the majority are metal-poor stars with [Fe/H] $< -1.0$, which appear to follow the low-$\alpha$ sequence of a typical dwarf galaxy. However, we are not able to confirm that they follow the same track of [Mg/Fe] versus [Fe/H] from the available abundance data. Further high-resolution spectroscopic studies are crucial to verify the hypothesis that they are from the same merger event.

\subsection{The Helmi Stream or S2}
\label{subsec:helmi}

This is one of the earliest pieces of halo substructure to be found kinematically. \citet{helmi99} first discovered a stream composed of eight stars in Hipparcos data. When this was supplemented with an extended dataset of metal-poor halo stars \citep{chiba00}, the stream was identified as having 12 members, separated into 3 stars with positive $v_z$ and 9 stars with negative $v_z$. This appeared to be a vindication of the power of searches in angular momentum space, which associated two different sets of debris in velocity space to the same substructure. \cite{He08} later provided a simulation that showed a possible interpretation of the data as the partially phase-mixed debris of a heavily disrupted satellite. Depending on the relative location of observer to the cloud of debris, we may expect to observe the debris as two separate velocity groups.

Subsequently, \citet{myeong18a} re-discovered the stream in their search for substructures in velocity space using {\it Gaia} DR1 data, cross-matched with the SDSS spectroscopic survey. They labelled it S2, as its connection with the structure found by \citet{helmi99} was not immediately apparent. A total of 73 stars were found in a striking, well-defined, and kinematically cold stream with prominent negative $v_z$ motion, which could correspond to the negative $v_z$ portion of the simulation from \citet{He08}. S2 revealed a clear stream feature for the first time -- the substructure was extended along the direction of its streaming motion. They fitted a model to S2, and argued that its progenitor had a total mass of $\sim  10^9 M_\odot$ and a stellar mass of $\sim10^6 M_\odot$, making it comparable to a present-day dSph like Draco.  \citet{myeong18c} searched for substructures in action space, and S2 was once again recovered. Since the actions can be considered as a set of integrals of motion, a group of stars with positive $v_z$ motion (with comparable action variables) were also included as a potential members in addition to the prominent negative $v_z$ stream. However, the positive $v_z$ portion appears more diffuse, without clear features in configuration space. The metallicity distribution function of S2 is peaked at [Fe/H] $= -1.9$ with a dispersion of 0.23 dex, ranging from $-2.65$ to $-1.3$ \citep{myeong18c}. 

After {\it Gaia} DR2, \citet{kopp18b} used its cross-match with APOGEE, RAVE, and LAMOST to identify more possible members. They drew a rectangular selection box in the space ($L_z, L_\perp = \sqrt{L_x^2 + L_y^2})$ to claim nearly 600 members of the stream, although of course this includes some contaminants from other halo stars that enter the box. They argued that the metallicity distribution ranges from [Fe/H] = $-2.3$ to 0.0, and peaks at [Fe/H]= $-1.5$ \citep[see Fig. 10. of ][]{kopp18b}. The authors considered the stars with [Fe/H] $\gtrsim -0.5$ to be likely contaminants. Their N-body simulations favored a system with a stellar mass of $\sim 10^8 M_\odot$ accreted 5--8 Gyr ago. This is significantly larger than suggested by \citet{myeong18a}. In their picture, the progenitor is an substantial dwarf galaxy, contributing approximately 15\% of its mass in stars. 

Based on the velocity properties of the 38 member stars of DTG-3 (see Table~\ref{tab:sub}), it can be readily  confirmed as the analog of S2 \citep{myeong18a}, 
similar to the negative $v_z$ part of the simulation of \citet{He08}. On the other hand, DTG-1 (with 9 stars) is a more diffuse substructure with positive $v_z$ motion, similar to the positive $v_z$ group mentioned above. It has slightly higher energy and a larger radial motion than DTG-3, as can be noted from the plots in the second row of Fig.~\ref{fig:action}. The pole of its angular momentum vector is also reversed because of the opposite sign of $v_z$. There appear to be two possibilities. First, such differences could arise if DTG-1 was stripped at an earlier pericentric passage from the same progenitor as DTG-3. Alternatively, DTG-1 may have a different origin than DTG-3 entirely, in which case DTG-3 would be the residue of a less-massive dwarf. High-resolution spectroscopic data should allow us to verify which of these hypotheses is more probable. 

The metallicity distribution function of this substructure can be examined here by using the map of the APOGEE halo sample onto the trained neuron network. This is shown in Fig.~\ref{fig:vphi_ecc_alpha}, with the two possible portions of the stream represented by orange triangles and circles. From the metallicity-abundance plot, they appear to be mostly $\alpha$-poor, and the metallicity distribution is truncated at [Fe/H] $\lesssim -1.3$. The metal-rich star at [Fe/H] $\approx -0.6$ is a clear contaminant, judged from its [Mg/Fe] and its orbital rotation velocity $v_\phi$, which marks it as a member of the thick disk.

\subsection{Prograde Groups}
\label{subsec:prograde}

Compared with the rest of the groups, DTG-2 (brown) and DTG-19 (yellow) are very prograde. Both have moderate eccentricity ($e \approx 0.6$ -- 0.8), but lowish inclinations ($i \approx 30^\circ$). They are both populated with stars having inward and outward $v_R$. This is typical of highly disrupted substructures, in which there may be multiple, co-existing wraps. The rotational velocity of DTG-19 is $\approx80$ km s$^{-1}$, which is characteristic of the metal-weak thick disk. The two clumps of DTG-19 with opposite $v_R$ are centered at $\approx 140$ kms$^{-1}$ and $\approx -157$ kms$^{-1}$, respectively. DTG-2 has a larger $\langle v_{\phi}\rangle \approx150$ km s$^{-1}$, as compared to DTG-19, as well as larger radial action and higher orbital energy. Both groups are in the region close to the disk stars in the ($J_\phi, E$) space. In the upper row of Fig~\ref{fig:as_ecc_new}, as well as in the second row of Fig~\ref{fig:action}, we have plotted the substructure Candidate Group 14 from \citet{myeong18c} as a red polygon. The latter plots especially suggest that DTG-2 is likely the same substructure as Candidate Group 14. 

Although DTG-2 and DTG-19 have lower eccentricities than the $Gaia$ Sausage groups, they are still high compared to typical disk stars. It is difficult to trace their origins from dynamics only, thus we look for chemical abundance information via their associated members from APOGEE. We find three candidate members associated with DTG-2 and five with DTG-19. From Fig.~\ref{fig:as_ecc_new}, we see that these members have orbital eccentricities, around 0.6 -- 0.8. Two of the three DTG-2 members, and three of the five DTG-19 members, sit in the region associated with the thick-disk system in the abundance versus metallicity plot. The other three stars from these two groups are very metal poor, so their abundances will have larger uncertainties. Although the sample size of APOGEE members is small, the [Mg/Fe] values of DTG-2 and DTG-19 favor the scenario that they are associated with the splashed or heated-disk component arising from the $Gaia$ Sausage merger event~\citep{DiM18,Bel19}.

Prograde orbits with strong radial motion match the description of the recently claimed Nyx substructures from \citet{necib19b,necib19a}. These two streams have oppositely directed subgroups, centered on $v_R$ at $\approx156$ kms$^{-1}$ and $\approx -124$ kms$^{-1}$, respectively, with rotational velocities of 141 kms$^{-1}$ and 120 kms$^{-1}$.  The similarities between the velocities of the Nyx substructures with the two prograde DTGs suggest that they too may be aspects of the same Splashed Disk phenomenon. 

\subsection{Polar Groups}
\label{subsec:polar}

The remaining groups all have relatively polar orbits. They are roughly separated into three different sets in the third row of Fig.~\ref{fig:action}. The maroon group (DTG-11) has slightly higher $J_z$ and $E$, whereas the orange groups and the purple groups are similar, but have different sign of $J_{\phi}$. The orange groups (DTGs-8, 36, 44, 54 and 55) are slightly retrograde, while the purple groups (DTGs-15, 34, 35, 39, 43, 45 and 51) are slightly prograde.

From velocity space, we see that the DTG-11 has a prominent vertical motion, with $\langle v_z \rangle$ = 230 km s$^{-1}$ , but almost zero rotation. It has a very polar orbit, which moves almost in the ($x,z$) plane, perpendicular to the plane of the Galactic disk. The orbital inclination is about 90$^\circ$, and the eccentricity is low, $e\approx 0.3$. DTG-11 almost connects with the orange and purple sets. The latter two have opposite inclination angle offsets, $\approx 20^{\circ}$ compared to DTG-11, and extended ranges of eccentricity. They are less polar than DTG-11, as clearly shown in Fig.~\ref{fig:as_ecc_new}.

From the distribution of DTG-11 in action space, it probably is a associated substructure with one of the metal-poor Candidate Groups from \citet{myeong18c}, who found a number of polar groups with low mean metallicity, such as Candidate 10 with $\langle[$Fe/H$]\rangle \approx -2.0$. Notice that all of the polar DTGs overlap with the {\it Gaia} Sausage groups in the ($J_\phi, E$) space, but have much more polar orbits. They may nonetheless be associated with the $Gaia$ Sausage, but were perhaps stripped off earlier when its orbit was less eccentric. The simulations of \citet{Amorisco17} show that massive satellites can lose their angular momentum to dynamical friction and gradually become more radial. The earlier stripped material may therefore have significant angular momentum, even if the final merger event is almost head-on.

From the APOGEE data, we find two members for the orange groups and 12 members for the purple groups. For both sets, half of the members reside in the low-$\alpha$ sequence, while the other half are in the high-$\alpha$ sequence. Although the metal-poor purple set appears to follow the lower-$\alpha$ track compared to the $Gaia$ Sausage stars, more members with detailed abundance measurements are required to verify their origin. Similar to the retrograde groups, the $\alpha$-rich members are originated from the thick-disk system.

\begin{table*}[ht]
\centering
\caption{Dynamical Properties of Existing Substructures}\label{tab:sub}
\bgroup
\def\arraystretch{2}
	\begin{tabular}{|c|c|cccc|}
		\hline
Substructure ($n_{\rm sub}$) & Groups& Component &($\langle v_R \rangle$, $\langle v_{\phi}\rangle$,$\langle v_z\rangle$) & ($\langle J_R \rangle$, $\langle J_{\phi} \rangle$, $\langle J_z \rangle$)& $\langle E \rangle$ \\
&&& ($\sigma_{v_R}$, $\sigma_{v_{\phi}}$,$\sigma_{v_z}$) &  ($\sigma_{J_R}$, $\sigma_{J_{\phi}}$,$\sigma_{J_z}$)& $\sigma_{E}$\\

&&& (km s$^{-1}$)&(kpc km s$^{-1}$) & km$^2$s$^{-2}$\\
 \hline
 
Helmi? (9)& DTG-1&  & (4.5, 197.2, 244.3)&(1118.6, 1839.4, 1118.0) &$-$1.1$\times$10$^5$ \\
                       &&                   &(146.0, 62.6, 42.4) & (700.0, 532.5, 400.1)&9.6$\times$10$^3$ \\

Helmi (38)     & DTG-3& &(26.2, 157.1, $-$241.3) & (253.2, 1367.9, 1123.7)&$-$1.3$\times$10$^5$ \\
                 &     &                    &(78.9, 28.8, 27.2) &(100.9, 236.9, 277.9) & 3.6$\times$10$^3$\\
\hline
         &  DTG-7,9,12,13,14,16,17,18,20,  &&(2.1, $-$0.3, $-$8.7)&(715.2, $-$3.7, 155.3) & $-$1.7$\times$10$^5$\\
Sausage (490)&25,26,27,30,31,32,37,38,40, &&(136.6, 35.0, 72.3)&(229.7, 296.2, 141.5)&1.2$\times$10$^4$\\
         & 41,42,46,47,48,49,50,56,57 &  & &&\\

\hline

Sequoia (27)&DTG-4,5 &&($-$36.9, $-$273.9, $-$87.0)&(712.4, $-$5.1, 169.4)& $-$1.3$\times$10$^5$\\
&&&(138.2, 36.7, 65.0)&(534.4, 293.8, 119.9)& 1.2$\times$10$^4$\\
\hline

       &DTG-6,52,53& $v_z$ (\texttt{+}) &($-$25.3, $-$98.0, 168.0)&(195.7, $-$872.8, 769.3)& $-$1.5$\times$10$^5$ \\
Rg5 (33) &                   &&(89.1, 23.5, 39.6) &(117.4, 244.6, 237.9) & 1.0$\times$10$^4$\\
       &DTG-10,23& $v_z$ (\texttt{-}) &(21.5, $-$74.5, $-$183.4)&(118.1, $-$637.4, 956.8)& $-$1.4$\times$10$^5$\\
       &     &                   &(79.7, 41.3, 56.8) &(95.5, 352.8, 331.0) & 4.1$\times$10$^4$\\

\hline
\end{tabular}
\egroup
  \medskip
\end{table*}

\begin{table*}[ht]
\centering
\caption{Dynamical Properties of New Groups}\label{tab:new}
\bgroup
\def\arraystretch{1.5}
\begin{tabular}{|c|c|cccc|}
		\hline
New& Groups &  Component &($\langle v_R \rangle$, $\langle v_{\phi}\rangle$,$\langle v_z\rangle$) & ($\langle J_R \rangle$, $\langle J_{\phi} \rangle$, $\langle J_z \rangle$)& $\langle E \rangle$ \\
&&& ($\sigma_{v_R}$, $\sigma_{v_{\phi}}$,$\sigma_{v_z}$) &  ($\sigma_{J_R}$, $\sigma_{J_{\phi}}$,$\sigma_{J_z}$)& $\sigma_{E}$\\
&&& (km s$^{-1}$)&(kpc km s$^{-1}$) & km$^2$s$^{-2}$\\
 \hline
& DTG-21 &&
(7.8, $-$129.3, $-$31.4) & (241.1, $-$1114.2, 48.0) & $-$1.7$\times 10^{5}$ \\
&&& (69.2, 21.4, 22.2) & (79.9, 183.0, 32.7) & 4.9$\times 10^{3}$\\
& DTG-24 &&
(19.1, $-$148.2, $-$87.5) & (123.9, $-$1279.4, 150.4) & $-$1.6$\times 10^{5}$ \\
&&& (50.9, 17.1, 11.8) & (41.5, 87.3, 65.8) & 1.9$\times 10^{3}$ \\
& DTG-29 &&
($-$16.7, $-$115.2, 29.2) & (280.6, $-$931.8, 51.1) & $-$1.7$\times 10^{5}$ \\
Rg&&& (67.8, 32.4, 20.2) & (115.3, 233.7, 44.1) & 4.3$\times 10^{3}$ \\
\cline{2-6}
& DTG-22 & &
($-$13.4, $-$107.5, 95.6) & (234.0, $-$903.2, 195.7) & $-$1.7$\times 10^{5}$ \\
&&& (66.4, 16.9, 19.1) & (76.0, 138.8, 66.4) & 5.2$\times 10^{3}$ \\
& DTG-28 & &
($-$4.0, $-$106.1, $-$143.2) & (287.1, $-$889.5, 478.3) & $-$1.6$\times 10^{5}$ \\
&&& (115.8, 29.3, 30.3) & (138.7, 185.3, 128.2) & 3.5$\times 10^{3}$ \\
& DTG-33 & &
(19.7, $-$94.8, $-$96.2) & (277.2, $-$761.4, 272.0) & $-$1.7$\times 10^{5}$ \\
&&& (84.4, 18.2, 36.0) & (147.7, 121.8, 163.2) & 3.4$\times 10^{3}$ \\
\hline
                                 &     &  $v_R$ (\texttt{+}) &(221.2, 155.7, 139.7)&(1071.2, 1372.1, 295.0)&  $-$1.3$\times$10$^5$\\
&DTG-2 (Cand14)               &     & (26.2, 33.8, 52.3)& (195.9, 289.7, 216.7)& 3.6$\times$10$^3$\\

                                 &     &  $v_R$ (\texttt{-})&($-$244.1, 211.5, $-$22.1)&(998.4, 1741.8, 83.3)& $-$1.3$\times$10$^5$ \\
\cline{2-6}
       
Pg                        & &$v_R$  (\texttt{+})&(139.7, 74.4, 82.6) &(550.4, 646.0, 154.6)&$-$1.6$\times$10$^5$ \\
&  DTG-19&                    &(22.6, 15.7, 25.7) &(110.2, 99.6, 72.5) &3.4$\times$10$^3$ \\
                        &  & $v_R$  (\texttt{-})&($-$157.0, 78.6, 60.8) &(580.5, 671.4, 136.7)&$-$1.6$\times$10$^5$ \\
                        &     &                    & (22.3, 21.7, 25.9)& (100.1, 156.9, 79.5)&2.3$\times$10$^3$ \\

\hline
     
& DTG-11 (Cand10?)& &($-$47.9, 21.8, 229.2)&(102.3, 192.3, 1830.6)&$-$1.4$\times$10$^5$  \\
 &     &                    &(75.4, 19.2, 21.5)&(41.9, 160.1, 126.0)& 1.9$\times$10$^3$ \\                                
\cline{2-6}
     
      & DTG-8 &&
(94.6, -45.7, 200.2) & (861.6, -389.6, 1068.9) & -1.3$\times 10^{5}$ \\
&&& (187.7, 45.0, 39.2) & (153.5, 370.5, 182.9) & 4.2$\times 10^{3}$ \\
& DTG-44 &&
(-3.2, -17.8, 170.4) & (167.2, -152.1, 1093.3) & -1.6$\times 10^{5}$ \\
&&& (69.9, 20.4, 18.6) & (57.4, 170.4, 133.1) & 1.0$\times 10^{3}$ \\
& DTG-54 &&
(-38.0, -47.4, 131.2) & (342.2, -401.5, 536.0) & -1.7$\times 10^{5}$ \\
&&& (86.5, 10.8, 11.5) & (82.0, 96.0, 146.1) & 2.0$\times 10^{3}$ \\
& DTG-36 &&
(9.2, -35.0, -143.7) & (310.2, -308.0, 785.9) & -1.6$\times 10^{5}$ \\
&&& (79.2, 12.8, 22.6) & (93.0, 102.7, 76.9) & 3.6$\times 10^{3}$ \\

\cline{2-6}
& DTG-39 &&
(15.3, 32.6, 135.2) & (330.7, 277.0, 567.5) & -1.7$\times 10^{5}$ \\
&&& (43.8, 11.0, 16.3) & (67.7, 104.5, 148.1) & 1.6$\times 10^{3}$ \\
Polar& DTG-43 &&
(-10.3, 59.4, 155.8) & (220.6, 489.7, 682.4) & -1.6$\times 10^{5}$ \\
&&& (81.6, 12.9, 15.7) & (88.3, 96.1, 82.9) & 3.0$\times 10^{3}$ \\
& DTG-15 &&
(12.1, 35.7, -185.5) & (575.7, 289.5, 1012.4) & -1.5$\times 10^{5}$ \\
&&& (174.8, 19.4, 44.8) & (164.5, 157.8, 289.9) & 3.1$\times 10^{3}$ \\
& DTG-34 &&
(-12.3, 93.3, -176.3) & (113.8, 736.7, 707.2) & -1.6$\times 10^{5}$ \\
&&& (49.0, 9.9, 9.1) & (77.1, 60.1, 97.7) & 2.2$\times 10^{3}$ \\
& DTG-35 &&
(30.5, 37.0, -160.5) & (216.6, 331.8, 952.0) & -1.6$\times 10^{5}$ \\
&&& (69.5, 11.7, 22.0) & (95.2, 105.2, 110.6) & 2.8$\times 10^{3}$ \\
& DTG-45 &&
(-13.6, 40.3, -113.0) & (430.5, 330.1, 348.1) & -1.7$\times 10^{5}$ \\
&&& (85.3, 17.2, 23.1) & (80.0, 129.7, 172.5) & 2.5$\times 10^{3}$ \\
& DTG-51 &&
(-9.1, 64.9, -138.6) & (301.0, 524.4, 493.1) & -1.6$\times 10^{5}$ \\
&&& (94.4, 6.3, 30.3) & (110.8, 43.8, 124.3) & 1.4$\times 10^{3}$ \\
\hline

\hline
\end{tabular}
\egroup
  \medskip
\end{table*}

\section{Early Nucleosynthesis Signatures}
\label{sec:chem}

We now search for $r$-process-enhanced and CEMP stars that may be dynamically associated with the VMP groups. We recall from Sec~\ref{sec:method} that the confidence of a given group member is defined as the probability of the Monte Carlo realizations of that member being associated with the group. The properties of the chemically peculiar members associated to different groups are listed in Table~\ref{tab:chem}.

\subsection{Association with r-Process-Enhanced Stars}

We find that four $r$-II stars in total are associated with two of the DTGs. CS~31082-001 and J2357-0052 are associated with DTG-10. We plot them by star symbols in dynamical space; they are seen to be well within the box of the Rg5 group from \cite{myeong18c} in Fig.~\ref{fig:as_ecc_old} and ~\ref{fig:action}. Both stars have very low metallicities, [Fe/H] = $-$2.78 and $-$3.36, respectively, and stand out compared to other $r$-II stars. First, both stars have an extreme enhancement of $r$-process elements, with [Eu/Fe] = +1.65 and +1.92, respectively. These exceed the nominal criteria of $\mathrm{[Eu/Fe]}\geqslant +1.0$ required for $r$-II stars \citep{beers2005}. In fact, among extremely metal-poor stars ($\mathrm{[Fe/H]<-3.0}$), SDSS~J2357-0052 is currently one of the most $r$-process-enhanced stars known \citep{Suda08}. CS~31082-001, on the other hand, is one of the best-studied $r$-II stars, with detailed abundances for 26 heavy elements beyond Sr, including the actinides Th and U. This star was the first example of a subclass of $r$-process-enhanced stars with Th and U abundances that are unexpectedly high, relative to Eu, referred to as ``actinide-boost" stars \citep{hill02}. It is not presently known if SDSS~J2357-0052 is also an actinide-boost star, as there is only a relatively high upper limit on [Th/Fe] ($< +2.74$; \citealt{aoki10}), and no measurement of U, available. Clearly, this star is of great interest for further study. Interestingly, based on the clustering of 35 highly $r$-process-enhanced stars with $\mathrm{[Eu/Fe]}\geqslant +0.7$ in the ($E$, $J_r$, $J_\phi$, $J_z$) space, these two stars belongs to the same Group C from \citet{roederer18}. This is consistent with our findings that they are dynamically associated with the same substructure, Rg5.

Using the SDSS-{\it Gaia} halo sample, \citet{myeong18c} show that Rg5 has very low mean metallicity ([Fe/H] = $-2.16$). According to the universal relation between stellar mass and metallicity for dwarf galaxies from \citet{kirby13}, the progenitor of Rg5 is likely a very low-mass dwarf galaxy with stellar mass of 10$^3$ -- 10$^6$M$_{\odot}$, such as an UFD. Our findings of the two $r$-II members with extremely low metallicities also favor the scenario of a very low-mass progenitor. Intriguingly, the absolute abundance of Eu in both the $r$-II stars in DTG-10 are identical (within $1\sigma$) with values of $\log\epsilon(\mathrm{Eu})= -0.92 \pm 0.19$ and $-0.76 \pm 0.11$. This is consistent with uniform enrichment of a low-mass dwarf galaxy by a single $r$-process event, similar to Reticulum II \citep{ji16,roederer16}.

Two additional $r$-II stars, J00405260-5122491 and 2MASS~J2256-0719, are found to be associated with DTG-38. Similar to DTG-10, the absolute abundances of Eu in the $r$-II stars are identical to each other. But unlike DTG-10, the metallicity  of the $r$-II stars are very similar ($\mathrm{[Fe/H]}\sim -2.2$), but much higher compared to DTG-10. Interestingly, the metallicity of the $r$-II stars in DTG-38 are similar to the typical metallicity of $r$-process-enhanced stars reported in classical dwarf spheroidals, such as Fornax \citep{letarte10,lemasle14} and Draco \citep{cohen09}. This is consistent with DTG-38 being the debris from a more massive system such as the $Gaia$ Sausage.  

\begin{table*}
\centering
\caption{Associated Chemically Peculiar Stars}\label{tab:chem}
\bgroup
\def\arraystretch{2}
\begin{tabular}{|c|c|cc|c|ccccc|}
		\hline
		    Substructure & Group  & Star ID & Type & Confidence & [Fe/H] &  [C/Fe]&  [Ba/Fe] & [Eu/Fe] & Reference\\
  \hline		   
\hline
     Helmi          & DTG-3 &HE~1135-0344         & CEMP-no  & 87 \% & $-$2.63 &  +1.03 &  \dots    &  \dots      & \citet{barklem05} \\ 

\hline
     Sausage        & DTG-13 & CS~22898-027     & CEMP-$i$   & 44 \% & $-$2.49 &  +2.11 &    +2.56     &  +1.92  & \citet{masseron12} \\

                    &        & SDSS~J0924+4059        &CEMP-$s$    & 41 \% & $-$2.51  &   +2.73 &   +1.86  & \dots &\citet{aoki08} \\
                  
\cline{2-10}
                    & DTG-18 & HE~1249-3121     & CEMP-no   & 26 \% & $-$3.23 &  +1.86 & \dots     & \dots    & \citet{barklem05} \\

\cline{2-10}
                    & DTG-38 & J00405260-5122491 &$r$-II &  100 \% &$-$2.11 &$-$0.04& $-$0.04& +0.86&\citet{hansen18}\\
                    &        & 2MASS~J2256-0719  &$r$-II & 28 \% &$-$2.26 &  +0.18 & +0.26 & +1.10 & \citet{sakari18}\\
                    &        & CS~22945-024 & CEMP-$s$ & 87 \% & $-$2.58 & +2.30 & +1.43 & +0.44 & \citet{roederer14}\\
                    &        & HE~0007-1832 & CEMP-no & 32 \% & $-$2.79 & +2.66&0.09&$<$ +1.75& \citet{cohen13} \\
\cline{2-10}
                    & DTG-41 & CS~22958-042 & CEMP-$s$ & 46 \% & $-$3.40 & +2.56 & $-$0.61 &$<$ +1.54& \citet{roederer14}\\

  \hline
  
    Sequoia         &  DTG-5    &CS~29514-007         & CEMP-no  &37 \%  &$-$2.83 & +0.89 &  $-$0.14 &   $< +1.74$   &\citet{roederer14}  \\

  \hline

    Rg5             & DTG-10 &SDSS~J2357-0052 &$r$-II    & 26 \% & $-$3.36  &  +0.43 & +1.08 & +1.92 &\citet{aoki10} \\ 
                    &      & CS~31082-001       & $r$-II & 25 \% &$-$2.90   & +0.29 &  +1.12     &  +1.62 & \citet{hill02} \\ 
                    & DTG-53 & HD~005223 & CEMP-$s$ & 28 \% & $-$2.11 & +1.58 & +1.88 & \dots &\citet{goswami06}\\ 

  \hline
    New Rg             & DTG-29 & CD-62:1346 & CEMP-$s$ & 42 \% & -1.59 & +0.86 & +1.58 & \dots &\citet{pereira12}\\
\hline
    New Rg            & DTG-28 & CS~29526-110 & CEMP-$s$ & 79 \% & $-$2.38 & +2.20 & +2.11 & +1.73 & \citet{aoki02}\\
    \hline
     New Rg               & DTG-33 & BD+04:2466 & CEMP-$s$ & 62 \% & $-$1.92 & +1.17 & +1.70 & \dots & \citet{pereira09}\\
\hline
    New Pg                 & DTG-19 & SDSS~J0212+0137 & CEMP & 65 \% & $-$3.57 & +2.28 & +0.16 & \dots & \citet{bonifacio15}\\ 
                    
  \hline
   New Polar                & DTG-45 & BD-01:2582 & CEMP-$s$ & 96 \% & $-$2.62 & +0.86 & +1.05 & +0.36 &\citet{roederer14}\\
                     &        & CS~22880-074 & CEMP-$s$ & 53 \% & $-$1.93 & +1.30 & +1.31 & +0.50&\citet{aoki02}\\
  \hline

	\end{tabular}
\egroup

  \vspace{2cm}
\end{table*}

\subsection{Association with CEMP Stars}

CEMP stars are important tracers of very early nucleosynthesis, and are more numerous than $r$-process-enhanced stars. They are frequently separated into sub-classses according to their neutron-capture element abundances \citep{beers2005}. CEMP-$s$ stars exhibit enhancements in their $s$-process elements, especially in Ba, and are frequently found to have radial-velocity variations \citep{starkenburg14,hansen16a}. Their high C and $s$-process element abundances can be understood as a result of the mass transfer from an evolved binary companion. CEMP-$i$ (formerly referred to as CEMP-$r/s$) stars have higher $r$-process-element abundances than CEMP-$s$ stars, but they are also likely to be the result of binary interaction. Their abundance pattern is well-described by neutron-capture reactions in an ``intermediate'' neutron density environment \citep{hampel2016}. On the other hand, CEMP-no stars show no enhancement in their neutron-capture elements, with a clearly lower binary frequency than CEMP-$s$ stars \citep{hansen16b}. Their C excess is generally attributed to nucleosynthesis pathways associated with the very first stars to be born in the universe \citep{iwamoto05,meynet06}. We are able to find nine CEMP stars associated with four existing substructures, and five with the newly identified ones. 

The $Gaia$ Sausage groups are associated with six CEMP stars of different sub-classes (CEMP-$i$, CEMP-$s$, and CEMP-no). DTG-13 hosts one CEMP-$s$ star and one CEMP-$i$ star. DTG-38, which is associated with two $r$-II stars, also has one CEMP-$s$ and one CEMP-no star. We find one CEMP-$s$ star in Rg5, and one in each of the five new DTGs (DTGs-29, 28, 33, 19, 45). Among these CEMP-$s$ stars, the one associated to DTG-29 is relatively metal-rich ([Fe/H] = $-$1.59).

Although CEMP-$s$ stars constitute a substantial subgroup in the stellar halo \citep{yoon19}, only one is known in all of the dwarf galaxies \citep{frebel14}. No CEMP-$i$ stars have yet been found in dwarfs. Our findings encourage the study of the birthplace of these heavy-element enriched CEMP stars in the stellar halo by tracing their dynamical origins.  Overall, all the CEMP-no associated in our study are found in substructures from classical dwarf galaxies. The Helmi Stream and Sequoia have one CEMP-no star each, and the $Gaia$ Sausage has two CEMP-no stars. Three of these four CEMP-no stars  have [Fe/H] $\approx -2.0$. Observationally speaking, we see more CEMP-no stars in this metallicity regime in classical dSph galaxies, compared to UFDs \citep[see Figure 1 of][]{yoon19}, though this may just reflect that fact that this is the \lq\lq metal-rich" tail of the MDF of UFDs.

\section{Discussion and Conclusions}
\label{sec:diss}

We have carried out a search for substructure in a catalog of 3364 very metal-poor (VMP) stars with halo kinematics extracted from LAMOST DR3. We used an neural-network algorithm based on Self-Organizing Map, \textsc{StarGO}, to identify groups of stars with similar dynamics in the input space of energy and angular momentum. The advantage of VMP stars is that they preferentially originate from the ultra-faint and other dwarf galaxies that are the building blocks of the stellar halo. The algorithm used here is different than previous works~\citep[e.g.,][]{myeong18a,myeong18c}, which typically build a data-driven smooth model against which to identify substructure as residuals. We identified 57 dynamically tagged groups (DTGs) comprising 972 member stars in total, using \textsc{StarGO} applied to a ``cleaned" version of the DR3 VMP catalog. Reassuringly, even though the algorithm and the dataset are very different, we are able to recover all of the known significant substructures in the nearby stellar halo, including the $Gaia$ Sausage~\citep{belokurov18}, Sequoia~\citep{myeong19}, the Helmi Stream \citep{helmi99}, the Splashed Disk \citep{DiM18,Bel19}, and some of the substructures found by \citet{myeong18c}, particularly Rg5. This helps build confidence in the reality of the substructures found both here and elsewhere.

The VMP stars potentially provide a great deal of information about the assembly history of the Milky Way. In a massive dwarf galaxy, such as the $Gaia$ Sausage progenitor, the VMP stars constitute a small fraction of its entire stellar population. Nevertheless, a massive progenitor contributes a significant portion of the VMP halo stars, because the number of the VMP stars scales with progenitor mass. More importantly, the clustering signatures are more pronounced in the VMP halo stars as compared to the full halo sample. The latter have a much larger smooth-halo population, dominated by stars with $-2 \la$ [Fe/H] $\la -1$ from a few massive merger events. 

Although our DTGs are identified in the VMP catalog, an attractive feature of \textsc{StarGO} is that we can match other catalogs onto the fully trained neuron map. This means that we can search for chemically peculiar stars associated with our substructures. Furthermore, we can take advantage of the precise abundances in APOGEE~\citep{Ab18} to probe the chemistry of our substructures as well. This considerably extends the power and applicability of the algorithm.

The {\it Gaia} Sausage is recovered as multiple groupings of VMP stars, all of which have characteristically high eccentricities ($e\ga$ 0.7). Examined in the plane of metallicity versus abundance using APOGEE data, these groups divide into two sequences. The first is a low-$\alpha$ track with the hint of a knee at [Fe/H] = $-1.3$, which is typical of massive dwarf galaxies like the $Gaia$ Sausage progenitor. The second is a high-$\alpha$ track with chemistry akin to thick-disk stars, despite their high eccentricity. This component is the Splashed Disk or Plume, identified in \citet{DiM18} and \citet{Bel19}. The impact of the $Gaia$ Sausage on the proto-disk of the Milky Way excited these stars onto high-eccentricity orbits. Although the existence of some Splashed Disk stars is expected, it is surprising how much of the high-eccentricity halo lies on the high-$\alpha$ track. 

One of the $Gaia$ Sausage debris is found to be associated with two $r$-II with similar but slightly higher metallicity that are typical of $r$-process enhanced stars found in classical dwarf spheroidals. This is consistent with the fact that the progenitor of the $Gaia$ Sausage was massive. The enhancement of $r$-process elements in such massive systems is expected to have large variations that are caused due to inhomogeneous mixing in their relatively large gas reservoirs (as well as mergers with lower-mass galaxies) resulting in both $r$-II and $r$-I stars. Thus, it is very likely that some $r$-I stars are originally from the $Gaia$ Sausage progenitor, but the dynamical associations are not sufficiently strong for our method to assign confident membership. 

The Sequoia is retrieved as two groups of VMP stars (DTG-4 and DTG-5), which are very strongly counter-rotating, and have chemistry consistent with a dwarf galaxy. Another retrograde substructure conspicuous in the VMP population is Rg5, first identified by \citet{myeong18c}. Interestingly, two highly $r$-process-enhanced stars with almost the same Eu abundances and very low metallicities, are dynamically associated with DTG-10 confidently. From observations of surviving UFDs such as Reticulum II, we know that these $r$-process-enhanced stars can be produced in low-mass dwarf galaxies, if they are enriched by prolific neutron-capture events~\citep[e.g.,][]{ji16}. Ultimately, Rg5 needs further study using high-resolution spectroscopy for the other group members to constraining the mass of the progenitor and to understand its early nucleosynthetic enrichment.

The Helmi Stream is seen in VMP stars potentially as two dynamically tagged groups, DTG-1 and DTG-3, which correspond to the (possible) positive $v_z$ portion and the negative $v_z$ portion, respectively. It is not entirely clear that DTG-1 and DTG-3 are parts of a single substructure, as their dynamical properties are similar, but differ in detail. DTG-3 is also much more populous, with 38 stars, than DTG-1, with just 9 stars.
\citet{He08} and \citet{kopp18b} interpret the Helmi Stream as the highly phase-mixed residue of a large dwarf galaxy which contributed 15 \% of the stellar mass of the Milky Way halo. The metallicity distribution function of the selected member stars in \citet{kopp18b} ranges from [Fe/H] = $-2.34$ extending to [Fe/H] = 0.0 \citep[see Fig. 10. of ][]{kopp18b}; the authors assumed the stars with [Fe/H] $\gtrsim -0.5$ are likely contaminants. Such a wide metallicity range requires a substantial progenitor. On the other hand, the APOGEE stars mapped on to DTG-1 and DTG-3 show a truncation at [Fe/H] $\la -1.3$ for the chemically cross-checked members, while the more metal-rich star appears to have a thick-disk origin to account for its high $\alpha$-element abundance. It is possible that the extended range of metallicity is partially affected by contamination from thick-disk stars.

In this work, we also identified 20 new groups. Six of them have retrograde orbits similar to Sequoia, but lower orbital energy. They are possibly inner-halo debris from the same merger event as Sequoia. We also find two very prograde substructures, with properties comparable to the present-day thick disk. Based on the APOGEE stars mapped onto them, both are $\alpha$-rich, but have high eccentricities compared to typical disk stars. They may be associated with the splashed or heated disk component generated by the $Gaia$ Sausage merger event~\citep{DiM18,Bel19}. The remaining substructures all have relatively polar orbits. While their orbital eccentricities are lower than typical $Gaia$ Sausage stars, they may nonetheless have a common origin. It is possible that they may be debris that was stripped earlier, before the orbit of the $Gaia$ Sausage progenitor became radialized.

In contrast to the rarity of $r$-process-enhanced stars, about 20$\%$ of VMP stars are also CEMP stars \citep{placco14}. We found that four known substructures (the $Gaia$ Sausage, Sequoia, the Helmi Stream, and Rg5) have CEMP members, as well as seven of the newly identified DTGs. The $Gaia$ Sausage has six associated CEMP stars, with a large diversity (CEMP-no, CEMP-$s$, and CEMP-$i$). The absolute amount of carbon in the members in each substructure could inform us as to the early chemical enrichment history of its birth place. 

This paper has taken the first steps in finding dynamical associations between disrupted dwarf galaxies and chemically peculiar stars, which is crucial for studying the birth environments of these accreted objects. This helps us understand the nucleosynthetic events and chemical evolution in UFDs and other dwarf galaxies in the early universe, by recovering the detailed abundance patterns for groups of stars from the same progenitor dwarf galaxy. The great advantage of studying debris in the nearby halo is that high-resolution spectroscopic observations are much more readily obtainable, so we expect this field of activity to have a rich future in the coming decades.

\bibliography{ms}
\bibliographystyle{apj}

\vfill\eject

\section*{Appendix}

In the original version of the LAMOST DR3 VMP catalog, published by \citet{li18}, it was recognized by a number of us that a substantial fraction (on the order of 20\%) of the stars included in the catalog did not, in fact, have spectra that suggested classification as VMP stars, for a variety of reasons. In order for us to employ an improved version of this catalog for the present analysis, spectra from the entire original version was inspected visually (by Beers), and the stars that were apparently problematic were noted.  We then ran the spectral data through a LAMOST version of the SSPP (which uses the flux-calibrated spectra), and another inspection was then carried out.  There still appeared to be a substantial number of problematic spectra, which were noted.  Because it was sometimes difficult to recognize which spectra are affected by poor flux calibration, we decided to run  all of them through a version of the n-SSPP, which does not use the flux calibration, but relies instead on a user-specified estimate of the appropriate $(g-r)_0$ colors (based on either the photometry in the original catalog, external photometry estimates, or, in cases where such information was not available, or suspect due to high reddening, estimates of the color derived from Balmer-line strengths).  

In the final cleaned version of the VMP table (which includes stars validated to have [Fe/H] $\le -1.8$, presented here in stub format, but available in full in the online version of this paper), the listed $g_0$ and $(g-r)_0$ come from the final adopted photometric estimates.  Estimates of $T_{\rm eff}$. log g, and [Fe/H] from the n-SSPP are supplied, along with estimates of [C/Fe] and [$\alpha$/Fe], where possible, are also provided.  The columns labelled "Det" in the table indicate whether the estimate of [C/Fe] or [$\alpha$/Fe] are considered as detections (D), upper limits (U), lower limits (L), or non-detections (N).

\begin{table*}
\centering
\caption{Cleaned LAMOST DR3 Very Metal-Poor Star Catalog}
\bgroup
\def\arraystretch{1.5}
\begin{tabular}{lcccccccccccccc}
\hline
\hline
~~~~~~~Star Name~~~~~~~ & $g_0$  & $(g-r)_0$  & $T_{\rm eff}$ & Error &log g&Error&[Fe/H]&Error&[C/Fe] &Error&Det&[$\alpha$/Fe]&Error &Det\\
&&&(K)&(K)&(cgs)&(dex)&&&(dex)&&(dex)&&(dex)\\

\hline
J000003.67+352146.0 & 12.974 &  0.418 &  5797 &    72 &  1.626 &  0.295 & -2.881 &  0.148 &  0.810 &  0.225  &D & \dots  & \dots  & \dots \\
J000040.57+103809.7 & 15.918 &  0.348 &  5534 &    90 &  4.052 &  0.168 & -1.875 &  0.150 &  0.051 &  0.126  &D &  0.361 &  0.039 & D \\
J000100.49+372356.8 & 16.130 &  \dots &  6049 &    50 &  1.420 &  0.598 & -2.590 &  0.168 &  1.912 &  0.123  &D & \dots  & \dots  & \dots \\
J000130.90+364056.4 & 16.369 &  0.402 &  6197 &    75 &  3.405 &  0.291 & -2.060 &  0.089 &  1.019 &  0.144  &D &  0.593 &  0.048 & D \\
J000150.63+113921.8 & 15.960 &  0.294 &  6251 &    64 &  3.349 &  0.203 & -1.995 &  0.150 &  0.728 &  0.123  &D & \dots  & \dots  & \dots \\
J000151.46+353921.8 & 13.438 &  0.521 &  6084 &    62 &  2.016 &  0.263 & -2.539 &  0.090 &  0.823 &  0.118  &U & \dots  & \dots  & \dots \\
J000220.64+413404.1 & 16.324 &  0.456 &  5435 &    71 &  3.153 &  0.158 & -1.986 &  0.126 &  0.202 &  0.091  &D &  0.408 &  0.042 & D \\
J000230.54+462026.1 & 12.811 &  0.393 &  6218 &    68 &  2.675 &  0.189 & -2.188 &  0.134 &  1.151 &  0.180  &D & \dots  & \dots  & \dots \\
J000232.81+263343.2 & 15.164 &  0.559 &  5222 &    47 &  2.945 &  0.241 & -2.284 &  0.157 &  0.536 &  0.097  &D &  0.438 &  0.053 & D \\
J000235.04+034337.5 & 15.116 &  0.220 &  6432 &    36 &  3.202 &  0.142 & -2.793 &  0.131 &  1.259 &  0.147  &D & \dots  & \dots  & \dots \\
\hline

\end{tabular}
\egroup
\end{table*}

\end{document}